\documentclass[aps,twocolumn,footinbib,floatfix]{revtex4-1}

\usepackage{CJK}
\usepackage{amsmath}
\usepackage{amsfonts}
\usepackage{soul}
\usepackage[caption=false,subrefformat=parens,labelformat=parens]{subfig}
\usepackage{color} 
\usepackage[table,dvipsnames]{xcolor}
\usepackage[colorlinks,linktocpage,bookmarks=false]{hyperref}

\newcommand\myshade{85}
\definecolor{myrulecolor}{RGB}{150,20,0}
\colorlet{mylinkcolor}{violet}
\colorlet{mycitecolor}{YellowOrange}
\colorlet{myurlcolor}{Aquamarine}
\hypersetup{
	linkcolor  = myurlcolor!\myshade!black,
	citecolor  = mycitecolor!\myshade!black,
	urlcolor   =  myrulecolor!\myshade!black,
}

\usepackage{graphicx}
\graphicspath{{FIG/}}
\usepackage{multirow}
\usepackage{braket}

\newcommand{\kb}{k_\text{B}}

\newcommand{\beq}{\begin{equation}}
\newcommand{\eeq}{\end{equation}}
\newcommand{\bea}{\begin{eqnarray}}
\newcommand{\eea}{\end{eqnarray}}

\makeindex

\begin{document} 
	\begin{CJK*}{UTF8}{gbsn} 
		\title{Hyperbolic Fracton Model, Subsystem Symmetry, and Holography II: The Dual Eight-Vertex Model
	}
		\author{Han Yan (闫寒)}
		
		\email{han.yan@oist.jp}
		\affiliation{Okinawa Institute of Science and Technology Graduate University, Onna-son, Okinawa 904-0495, Japan}
		\date{\today}
		
\begin{abstract}
The discovery of fracton states of matter
opens up an exciting, largely unexplored field of many-body physics.
Certain fracton states' similarity to gravity is an intriguing property.
In an earlier work \cite{Yan2018arXiv},
we have demonstrated that a simple
 fracton model in anti-de Sitter space satisfies 
several major holographic properties.
In this follow-up paper,
we study the eight-vertex model 
dual to the original model.
The dual model has the advantage of 
illuminating the mutual information and subsystem charges
pictorially,
which helps to reveal its connections to various other 
topics in the study of holography and  fracton phases.
At zero temperature,
the dual eight-vertex model 
is a discrete realization of the 
bit-thread model,
a powerful tool developed to visualize holography.
The bit-thread picture combined with subsystem charges can
give a quantitative account of the isometry between the bulk and the 
boundary at finite energy,
which is also a key issue for  holography.
The black hole microscopic degrees of 
freedom can be identified in this picture,
which turn out to be encoded non-locally 
on the horizon.
The eight-vertex model
proves to be a very helpful venue 
to improve our understanding of the hyperbolic fracton model
as a toy model of holography.

\end{abstract}
\maketitle
\end{CJK*}


\section{Introduction}
The recent discovery of fracton states of matter \cite{
ChamonPhysRevLett.94.040402,YoshidaPhysRevB.88.125122,BRAVYI2011839,Haah2011,Vijay2015,Vijay2016,Pretko2017a,Pretko2017b,Nandkishoreannurev}
is an exciting development in many-body physics.
These models feature 
exotic excitations with constrained mobility dubbed ``fractons'' , 
and
gauged or ungauged subsystem symmetries. 
The fracton topological orders
are also beyond our conventional knowledge
of topological orders.
The fracton states
present many new challenges,
including 
model building
\cite{Shirley2017,Slagle2019SciPost,Tian2018arXiv181202101T,YouYZ2019arXiv,
SongPhysRevB.99.155118,YouPhysRevB.98.035112},
experimental realizations \cite{SlaglePRB17,hseih17,Halasz2017,Ma2017,You2018arXiv,Yan2019arXiv,Benton2016},
proper classification scheme \cite{Shirley2018,Pai2019,Gromov2018arxiv,Shirly2019SciPost},
quantum-information application
\cite{
Schmitz2018,Kubica2018,He2018PRB,Ma2018a,Schmitz2018arXiv180910151S,Schmitz2019arxivES,PaiPhysRevX2019,ShirleySciPostPhys2019},
and its connection to other  areas of physics \cite{Pretko2017,Pretko2018PRL,PaiPhysRevB.97.235102,PretkoPhysRevLett20182,gromov19PRL,Yan2018arXiv}.

An intriguing aspect of fracton states of matter 
is their similarity to gravity \cite{Pretko2017,gromov19PRL,Slagle2019SciPost}.
The fracton excitations can be described as charges of
the generalized rank-2 U(1) gauge theories \cite{XuPRB06,rasmussen2016stable,Pretko2017a,Pretko2017b,Slagle2017-2,Prem2018arXiv180604687P,You2018arXiv180509800Y,Bulmash2018,Ma2018,Slage2018arXiv180700827S,ShirleyPhysRevX.8.031051,DevakulSciPostPhys.6.1.007},
where the electric and gauge fields take 
the form of symmetric matrices and have modified Gauss conservation laws.
These theories have been shown 
to exhibit behaviors similar to 
general relativity,
and are indeed the linearised limit 
of certain gravitational/elasticity theories \cite{Pretko2017,Pretko2018PRL,gromov19PRL}.

Along this line,
a very simple classical fracton toy model 
in anti-de Sitter space 
was shown to satisfy a few major holographic 
properties \cite{Yan2018arXiv}.
The holographic principle \cite{Hooft1974,Susskind1995} and anti-de Sitter/conformal field theory (AdS/CFT) correspondence \cite{Maldacena1999,Witten1998},
as a ground-breaking framework to demystify quantum gravity,
have been front line for the high energy theory community for a few decades \cite{gubser1998gauge,Aharony2000,Aharony2008,Klebanov2002,Hawking2005,Guica2009}.
It is also a powerful tool-set to understand strongly coupled systems \cite{Hartnoll,PiresAdSCFT2014,ZaanenHolographyCMT2015,NastaseStringCMT2017,Qi2013,Gu2016,Lee2016}.
In this context, the hyperbolic fracton model
satisfies the celebrated Ryu-Takayanagi formula \cite{Ryu2006,Ryu2006a},
and also has the correct subregion duality \cite{Raamsdonk2009arXiv}.
Its construction has a lot of similarities to the holographic toy models
built from tensor networks \cite{Swingle2012,Pastawski2015,Almheiri2015,Yang2016,Hayden2016,Qi2018,Harlow2017CMaP}.

This paper, as the second of the duology on the hyperbolic fracton model,
studies the dual eight-vertex model of the original model,
which has the advantage of visualizing the mutual information and subsystem fluxes.

It helps to address a few key unanswered questions following the initial discovery.
One question is whether the hyperbolic fracton model is
equivalent to any other known holographic models/theories.
This turns out to be true. 
The dual eight-vertex model
is a discrete realization of the  bit-thread model
\cite{Freedman2017CMaPh,Cui2018arXiv,Harper2018arXiv,Headrick2018CQG,Chen2018arXiv},
which was proposed as a very powerful framework to understand holography.
It treats the non-local ``flow of information'' instead 
of  local fields 
as the elementary physical quantity.
From this perspective
many holographic properties of 
entanglement entropy have an intuitive, pictorial derivation.

Another question is about holography
beyond the ground states.
This was not discussed much in the previous work.
Here equipped with the bit-thread picture
and the concept of subsystem charges,
a detailed analysis is presented.
We show that ``isometry'',
the requirement from holography that the boundary uniquely determines
the bulk,
is violated only by a small amount at low energy levels,
and all violating cases can be determined.

The bit-thread and subsystem charge language
also help us to identify the black hole microscopic degrees of freedom (dofs),
which is encoded non-locally on the horizon, and also the AdS boundary.
Intriguingly even though the black hole set-up is very primitive,
it yields qualitatively correct behavior of how 
a boundary observer can distinguish the microstates \cite{Bao2017PRD}. 

This work and Ref.\cite{Yan2018arXiv} form a 
relatively comprehensive investigation of the 
classical toy hyperbolic fracton model.
In the outlook, 
we discuss
future directions beyond this simple toy model, 
which could be an interesting program for condensed matter physics,
and hopefully provide some insights in high energy theory too.

This paper is arranged as follows:
Sec.~\ref{SEC_II_Fracton_model} briefly reviews the results from Ref.~\cite{Yan2018arXiv}; 
Sec.~\ref{SEC_3_Vertex_model_square} describes the dual eight-vertex model
on the Euclidean lattice and 
Sec.~\ref{SEC_4_Vertex_model_hyperbolic} 
on the hyperbolic lattice;

The first major result of this work,
Sec.~\ref{Sec_5_Bit_thread}, explains the eight-vertex model 
as a realization of the bit-thread model.
It is then utilized  to  derive results documented in the two following sections:
Sec.~\ref{Sec_6_Isometry} analyzes the isometry properties 
of the excited states;
Sec.~\ref{Sec_7_BH} describes the black hole microscopic degrees of freedom
in the model;

Finally Sec.~\ref{Sec_8_outlook} summarizes this paper and gives an outlook
of possible future directions.

\section{Brief Review of the Holographic Hyperbolic Fracton Model } \label{SEC_II_Fracton_model}

In this section we recapitulate the  classical hyperbolic fracton model 
and its holographic properties,
which are the main result of Ref.~\cite{Yan2018arXiv}.
Interested readers are recommended to refer to it
for more details.

\begin{figure}[ht]
	\centering
	\subfloat[Square Lattice\label{Fig_Frac_a_Square}]{\includegraphics[height=0.25\textwidth]{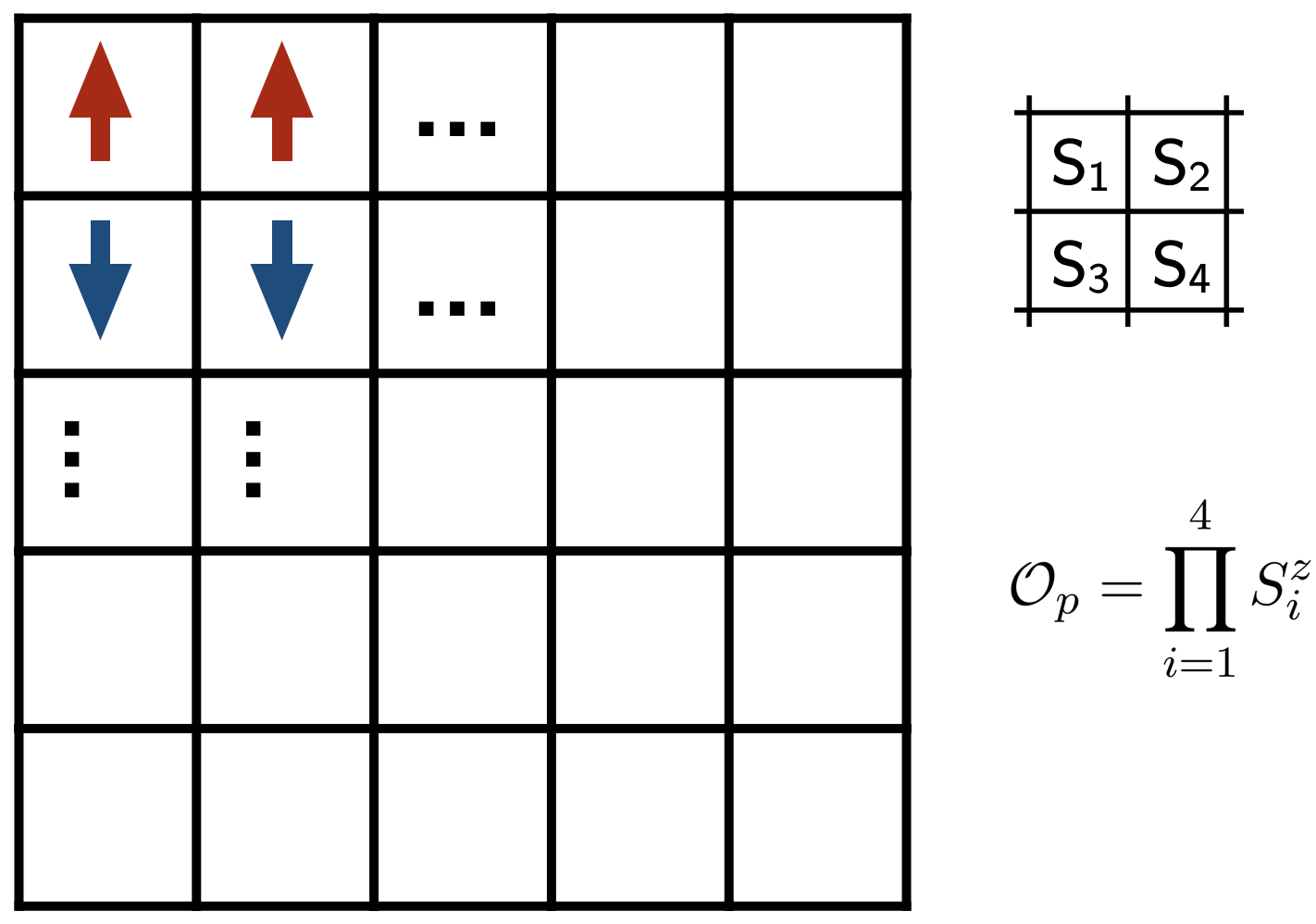}}\\
	\subfloat[Hyperbolic lattice\label{Fig_Frac_b_Hyper}]{\includegraphics[height=0.25\textwidth]{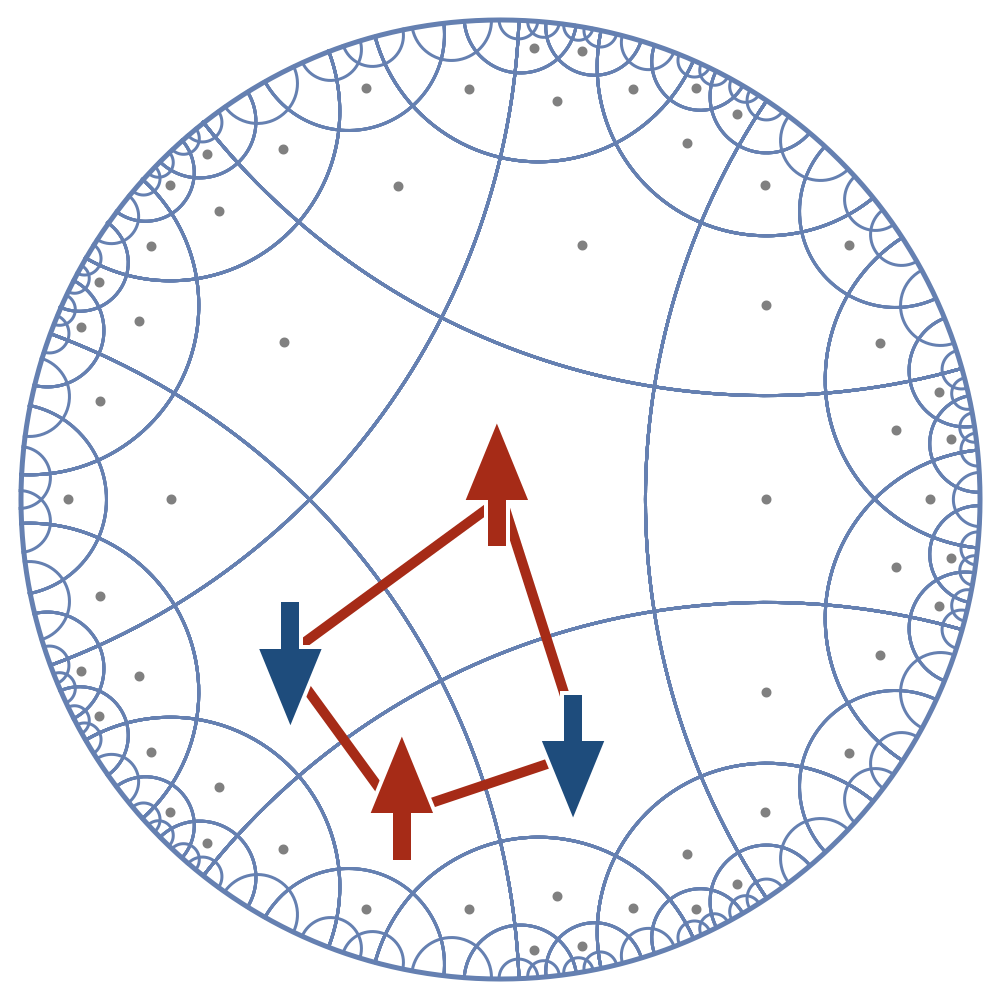}}
	\caption{The fracton model (Eq.~\eqref{Eqn_Ham_Fracton}) on the Euclidean and hyperbolic lattice.
		(a) The model defined on the Euclidean lattice.
		(b) The model defined on the hyperbolic lattice of (5,4) tessellation. Spins sit at the centers of the pentagons. The plaquette operator
		is defined on each cluster of the four corner-sharing pentagons as shown by the red box. 
	} 
	\label{Fig_Frac}
\end{figure}

The classical fracton model can be defined on 
both the Euclidean and hyperbolic (negatively curvatured, or AdS) lattice
based on uniform square and pentagon tessellations shown in Fig.~\ref{Fig_Frac_a_Square},~\ref{Fig_Frac_b_Hyper}.
In the later case, 
the hyperbolic lattice is obtained by the (5,4) tessellation, i.e., tiling the 2D AdS space with pentagons,
with four pentagon sharing every corner.
An Ising spin of value $S_i^z = \pm1$ is placed at the \textit{center} of each square in the Euclidean lattice or pentagon in the hyperbolic lattice.
The operator 
\begin{equation}\label{EQN_Op_Def}
\mathcal{O}_p=\prod_{i=1}^{4} S^z_i,
\end{equation}
is defined for each four-spin cluster, where $i$ runs over its four sites.
Such a  cluster on the hyperbolic lattice is shown by the red rectangle in Fig.~\ref{Fig_Frac_b_Hyper}.
The Hamiltonian for both models is
\begin{equation}\label{Eqn_Ham_Fracton}
\mathcal{H}_\textsf{spin} = -\sum_p \mathcal{O}_p		\;,
\end{equation}
where the sum runs over all four-spin clusters.

\begin{figure}[ht]
\centering
\subfloat[]{\includegraphics[height=0.22\textwidth]{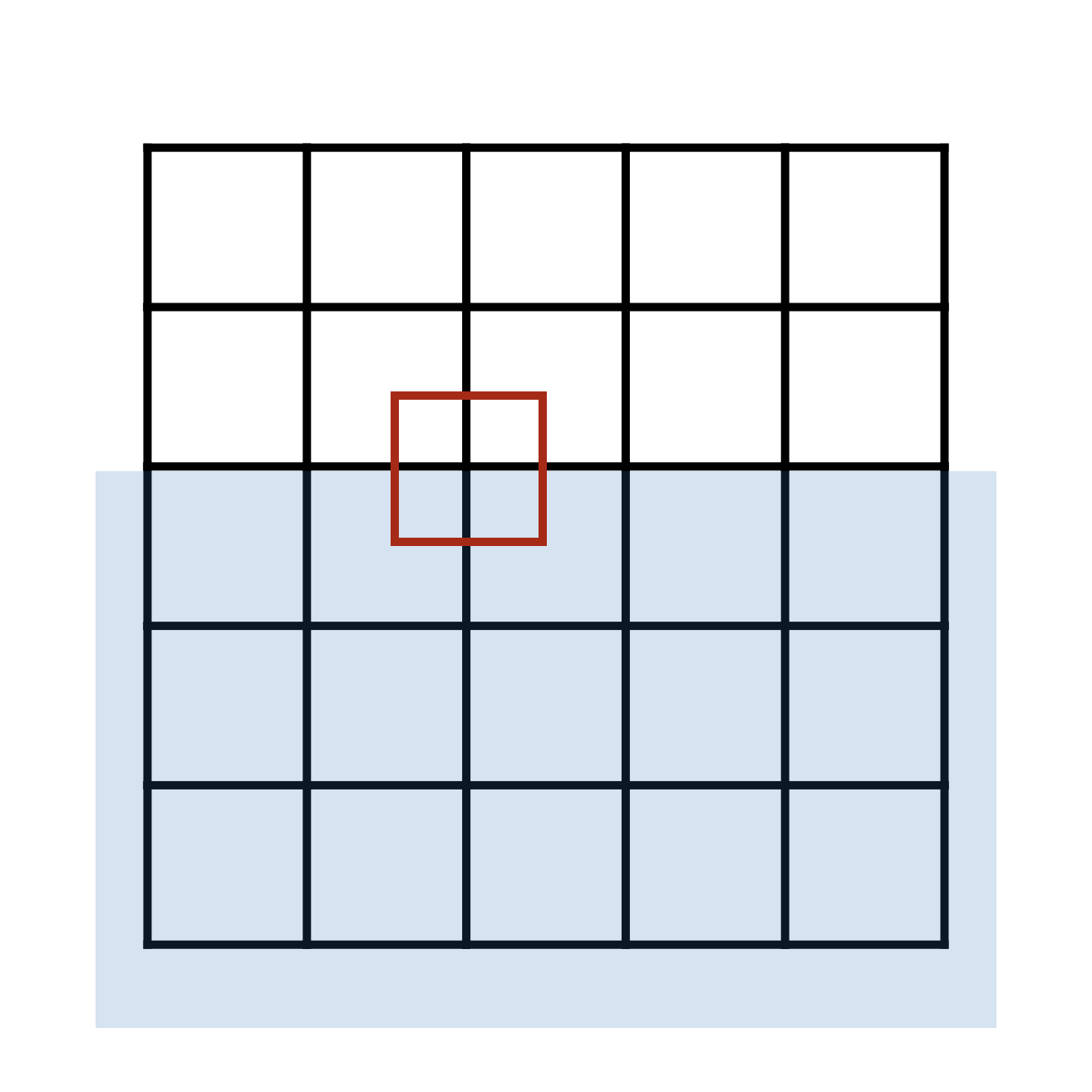}}\quad
\subfloat[]{\includegraphics[height=0.22\textwidth]{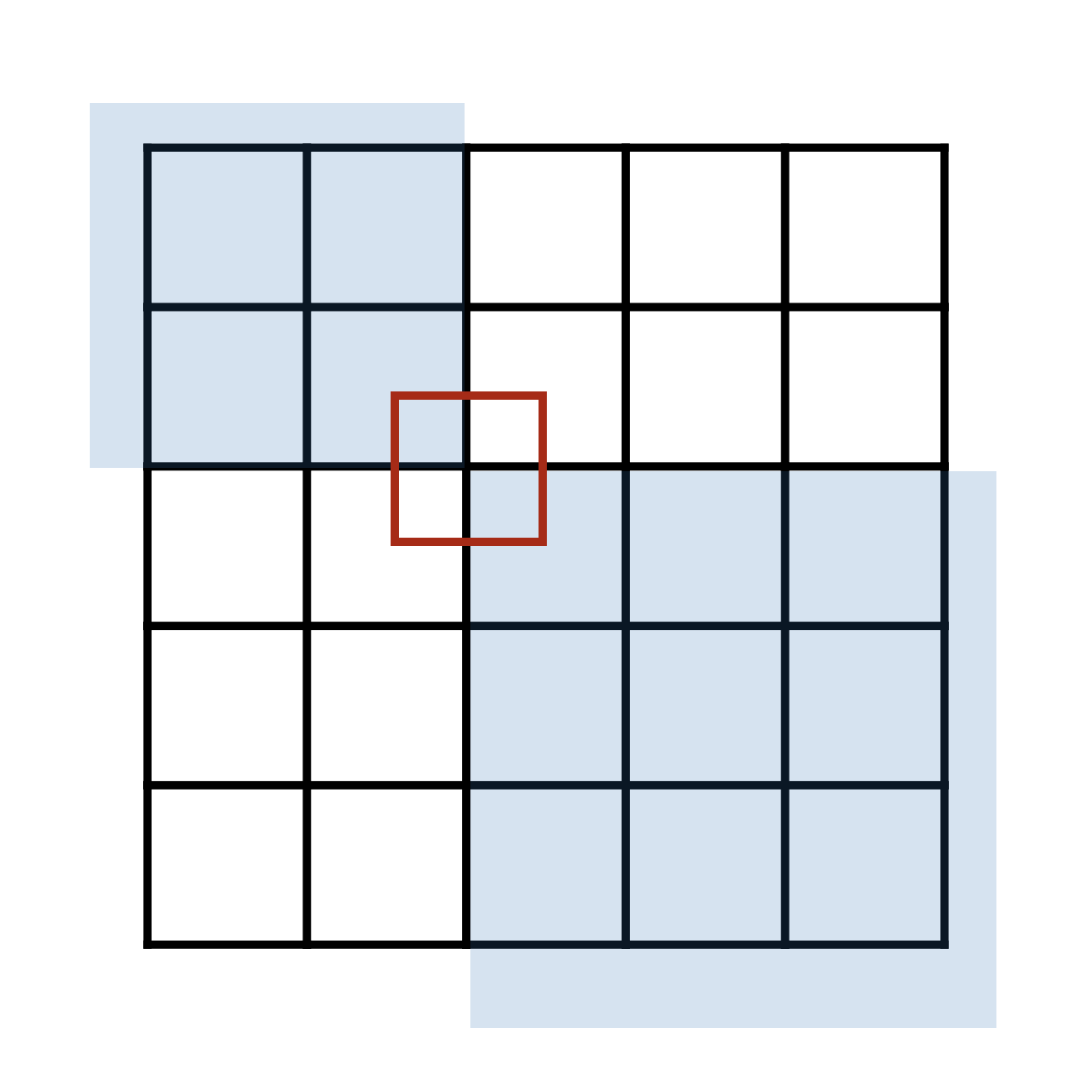}}
\\
\subfloat[]{\includegraphics[height=0.22\textwidth]{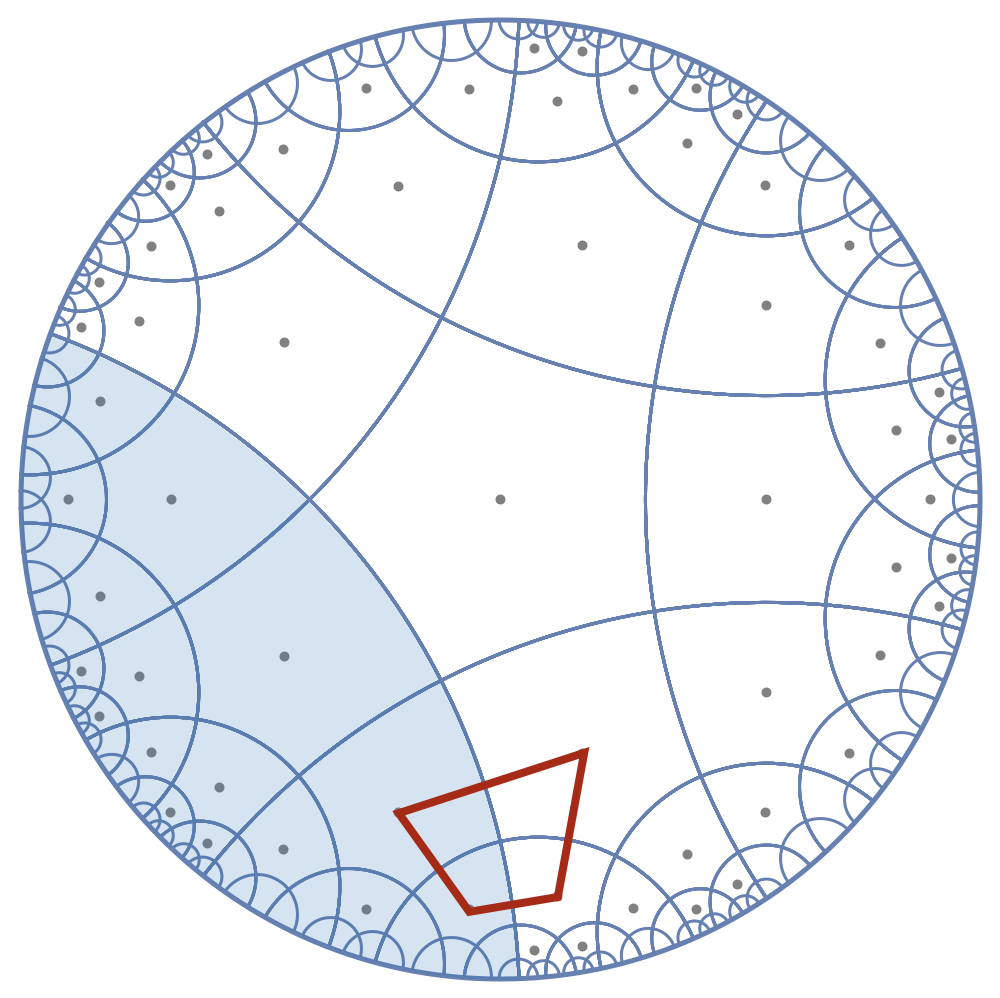}}\quad
\subfloat[]{\includegraphics[height=0.22\textwidth]{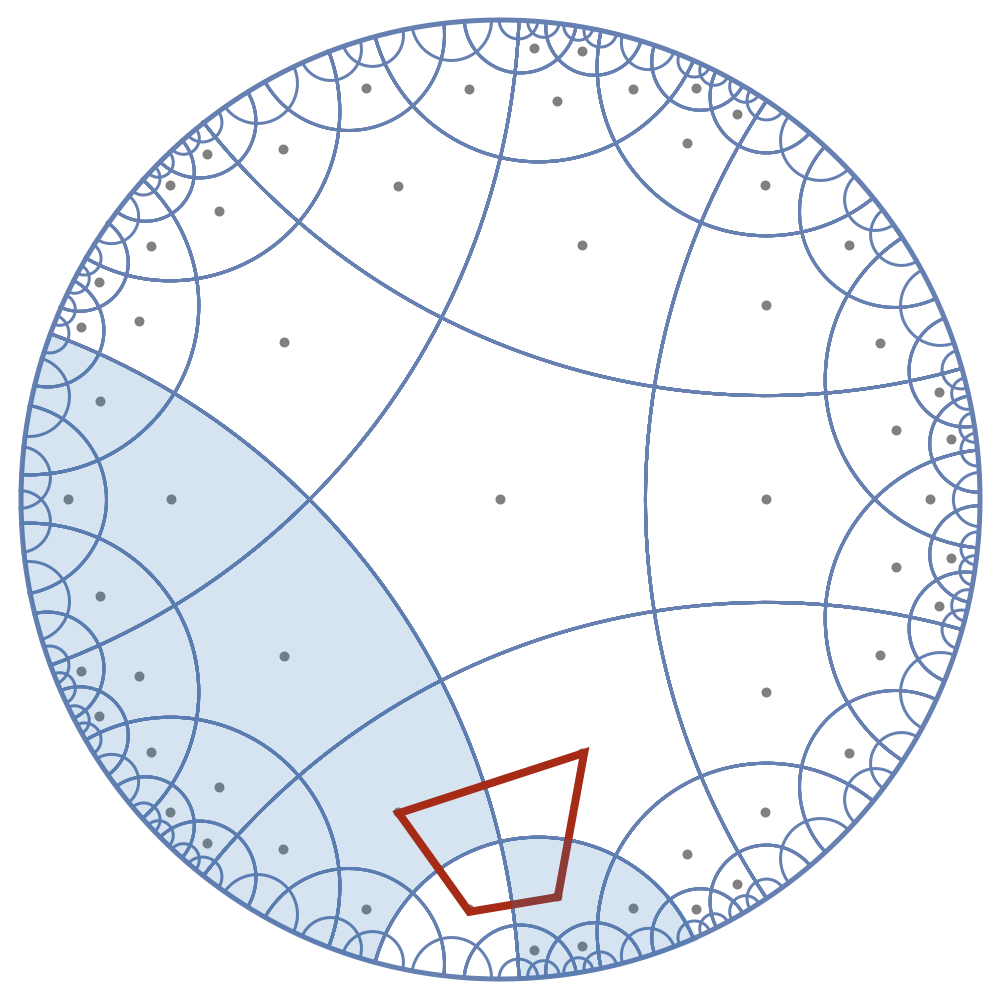}}
\caption{Ground state degeneracy of the fracton models on the Euclidean lattice (a, b) and hyperbolic lattice (c, d).
The blue region are the spins flipped from the original ground state.
Note that the four-spin clusters always have even number of spins flipped.
} 
	\label{Fig_Frac_GS}
\end{figure}

An essential property of these models is their subsystem symmetry.
Note that the pentagon's edges define some geodesics,
i.e.,
straight lines in $x-$ or $y-$ direction on the Euclidean lattice and
arcs intersecting the disk boundary perpendicularly
on the hyperbolic disk.
The energy of the system is
invariant under 
the operation of 
flipping all spins on either side of a chosen geodesic.
By starting from any given ground state and consecutively applying such operations for different geodesics,
all ground states can be explicitly constructed.
Thus, the ground state degeneracy is proportional to 
$2^\text{number of geodesics}$,
which is also  proportional to $2^\text{boundary size}$.
This feature is dubbed ``subsystem symmetry'' in literature \cite{Vijay2016,YouPhysRevB.98.035112,Shirly2019SciPost}.
It is
a symmetry in-between local and global,
and the origin of many exotic features of fracton models,
including the holographic ones.\\
%

Ref.~\cite{Yan2018arXiv} has demonstrated the following holographic properties of the hyperbolic fracton model.

\begin{figure}[ht]
	\centering
	\subfloat[\label{Fig_Fracton_property_1}]{\includegraphics[width=0.21\textwidth]{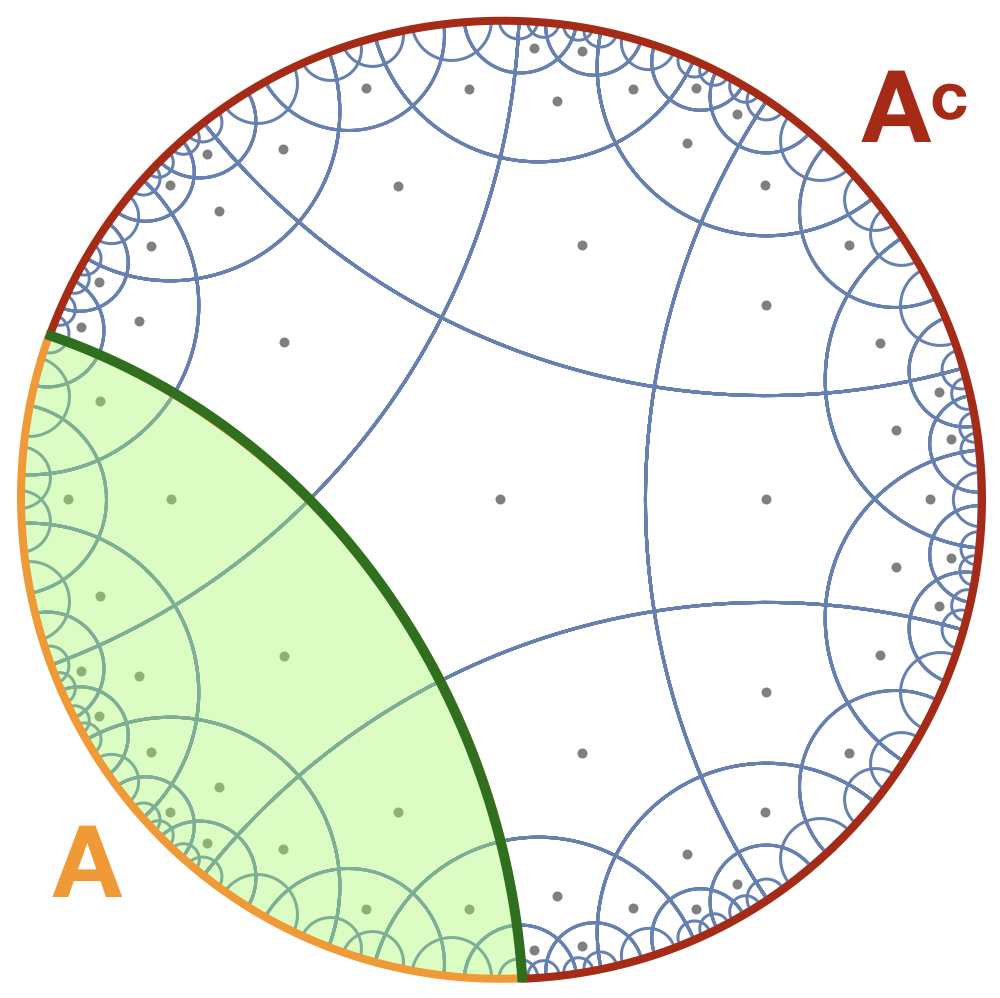}}\quad
	\subfloat[\label{Fig_Fracton_property_2}]{\includegraphics[width=0.21\textwidth]{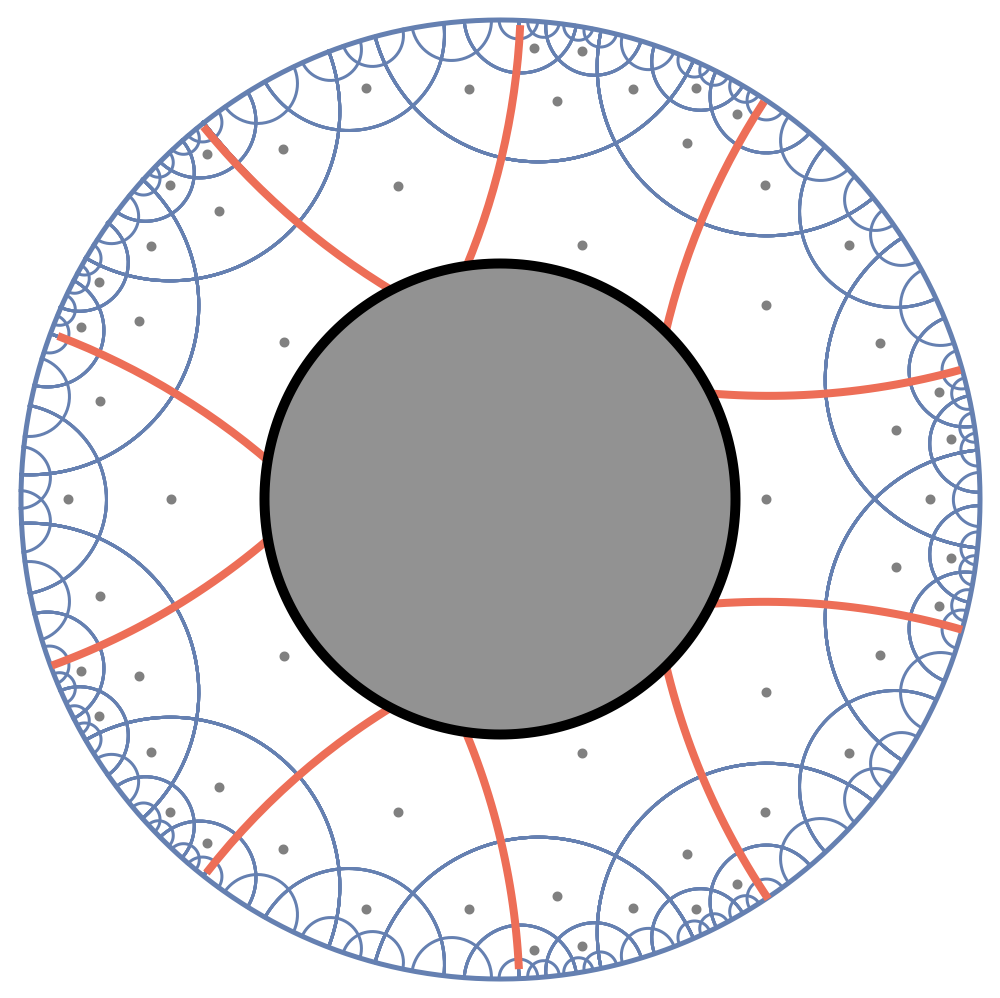}}\
	\caption{Holographic properties of the hyperbolic fracton model. 
		(a): Sub-region duality and RT formula for mutual information.
		Give the boundary spins on segment $A$ (orange),
		the reconstructible bulk at $T=0$ are those in the minimal convex wedge colored in
		light green.
		The mutual information between boundary bipartition $A$ and $A^c$ (red) is proportional to the 
		length of the minimal covering surface (dark green geodesic).
		(b): A black hole in the bulk. This naively defined black hole (dark gray region) has
		entropy proportional to its horizon area, or the number of orange geodesics.
	} 
	\label{Fig_Fracton_property}
\end{figure}

\noindent
{\bf Rindler reconstruction and subregion duality ---} 
For the hyperbolic fracton model defined by Eq.~\eqref{Eqn_Ham_Fracton}, 
given a spin configuration on a connected boundary segment,
the bulk spins in a specific region (Fig.~\ref{Fig_Fracton_property_1}) are determined unambiguously at zero temperature.
This region, dubbed  {\it minimal convex wedge},
agrees with the reconstructible entanglement wedge 
determined by Rindler reconstruction or subregion duality of holography \cite{Raamsdonk2009arXiv}.\\

\noindent 
{\bf  Ryu-Takayanagi formula for mutual information ---}
Given a bipartition of the boundary into two connected
segments $A$ and $A^c$,
their mutual information (the classical analog of entanglement entropy),
\begin{equation}
\I(A,A^c)  = S_A + S_{A^c} - S_{A \cup A^c},
\end{equation}
obeys the Ryu-Takayanagi formula \cite{Ryu2006,Ryu2006a}:
\begin{equation}\label{Eqn_RT}
\I(A,A^c) =\kb\log2\times |\gamma_A| \;.
\end{equation}
where $|\gamma_A|$ is the area of the minimal covering surface, 
or in this case the length  of the geodesic 
that separates $A$ and $A^c$ (Fig.~\ref{Fig_Fracton_property_1} ).\\

\noindent 
{\bf Black hole entropy  ---}
A very naively defined black hole in the system,
i.e., with a convex horizon but not changing the AdS geometry,
has entropy proportional to the area of its horizon (Fig.~\ref{Fig_Fracton_property_2}),
\begin{equation}\label{Eqn_BH_Entropy}
S_\mathsf{BH} =\frac{\kb\log2}{2}\times |\gamma_A| \;.
\end{equation}
which is consistent with the Hawking-Bekinstein black hole entropy  \cite{Hawking}.

The difference of the factor $2$ between Eq.~\eqref{Eqn_RT}
and Eq.~\eqref{Eqn_BH_Entropy} is consistent, since by definition
the mutual information is twice the entanglement entropy \cite{Yan2018arXiv}.

\section{Dual Eight Vertex Model on the Square Lattice} \label{SEC_3_Vertex_model_square}
The main results of this paper revolve around
a physically equivalent model of the hyperbolic fracton model ---
the dual eight-vertex model.
Formulated in the language of arrows and vertices,
it has the advantage of illuminating various connections
between the hyperbolic fracton model and other 
established results in fracton phases and holography.
In this section, 
we will describe the dual eight-vertex model,
and discuss how it works as a 
straightforward demonstration of  
fracton-elasticity duality \cite{Pretko2018PRL,PretkoPhysRevLett20182,gromov19PRL}
and subsystem charge \cite{Shirly2019SciPost}.

\begin{figure*}[ht]
	\begin{center}
		\begin{tabular}{c|c|c|c|c|c|c|c|c}
			\hline
			\multirow{2}{*}{vertex} & 1 & 2 & 3 & 4 & 5 & 6 & 7 & 8 \\
			& \includegraphics[width=0.07\textwidth]{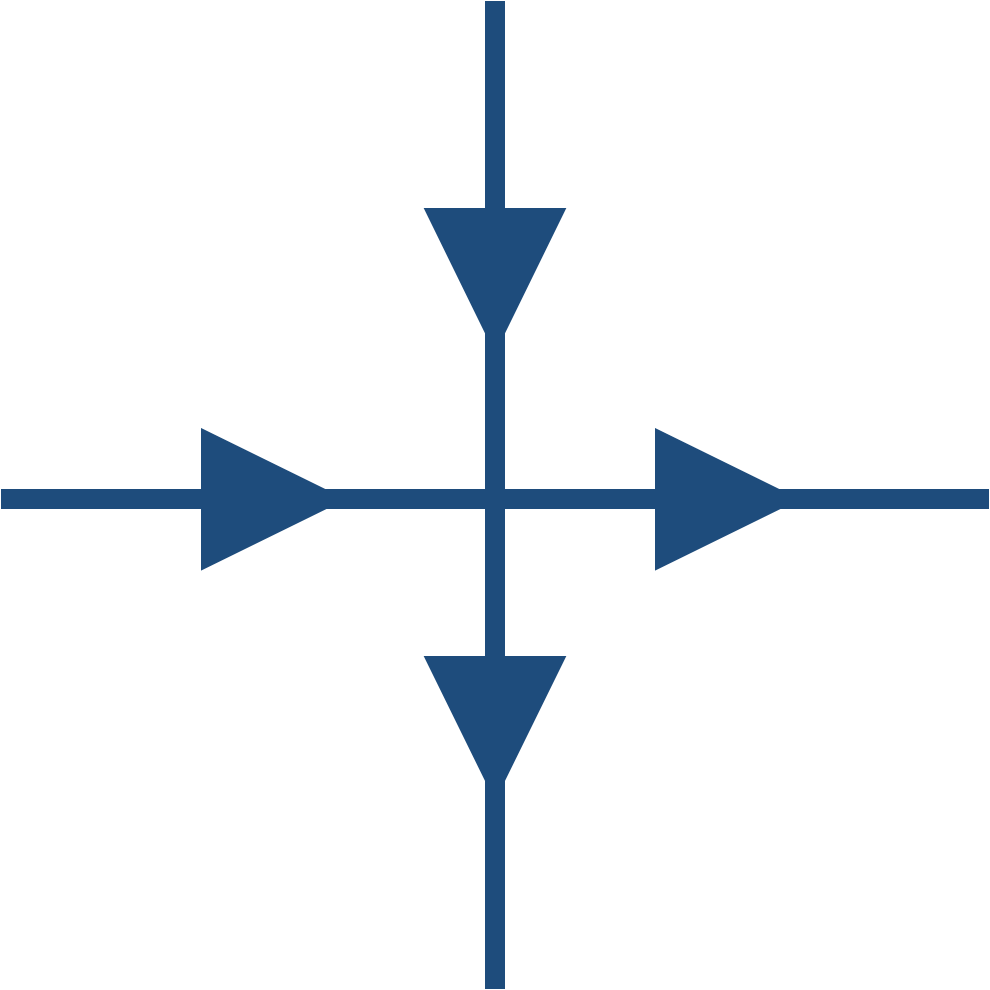} 
			& \includegraphics[width=0.07\textwidth]{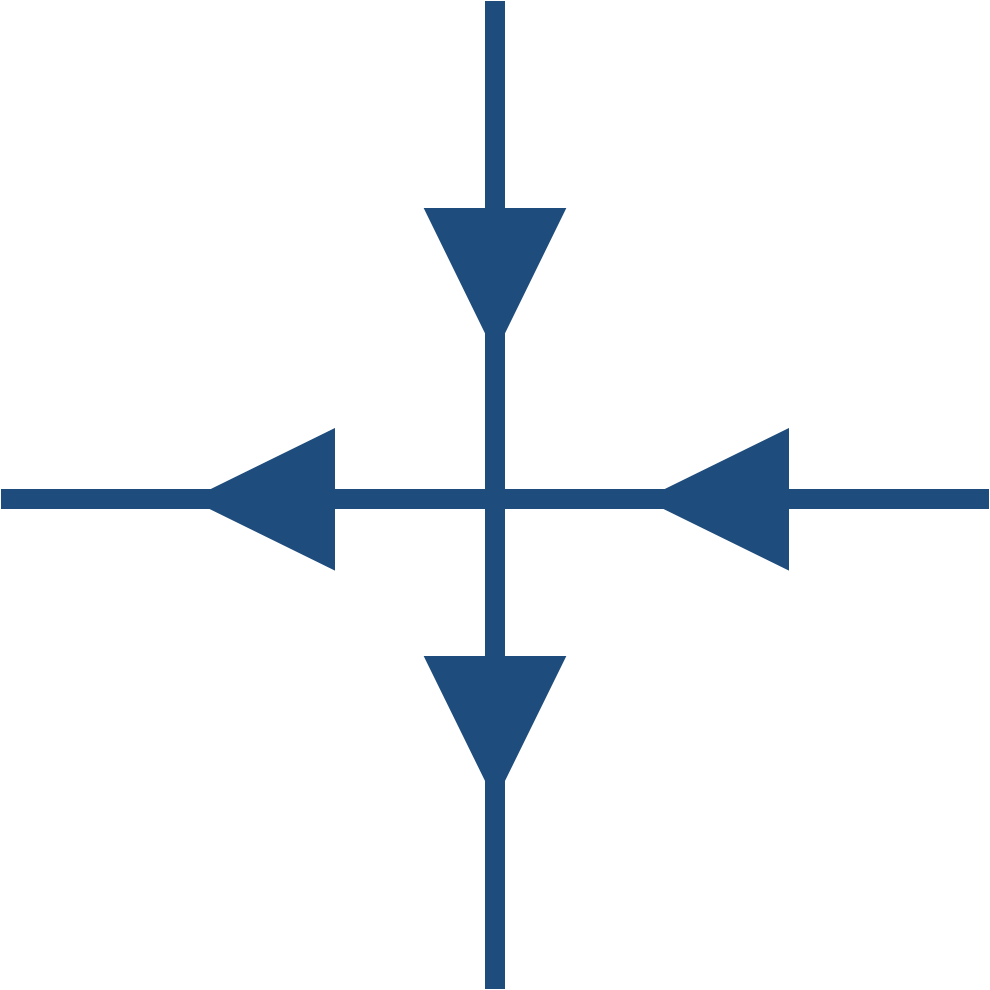} 
			& \includegraphics[width=0.07\textwidth]{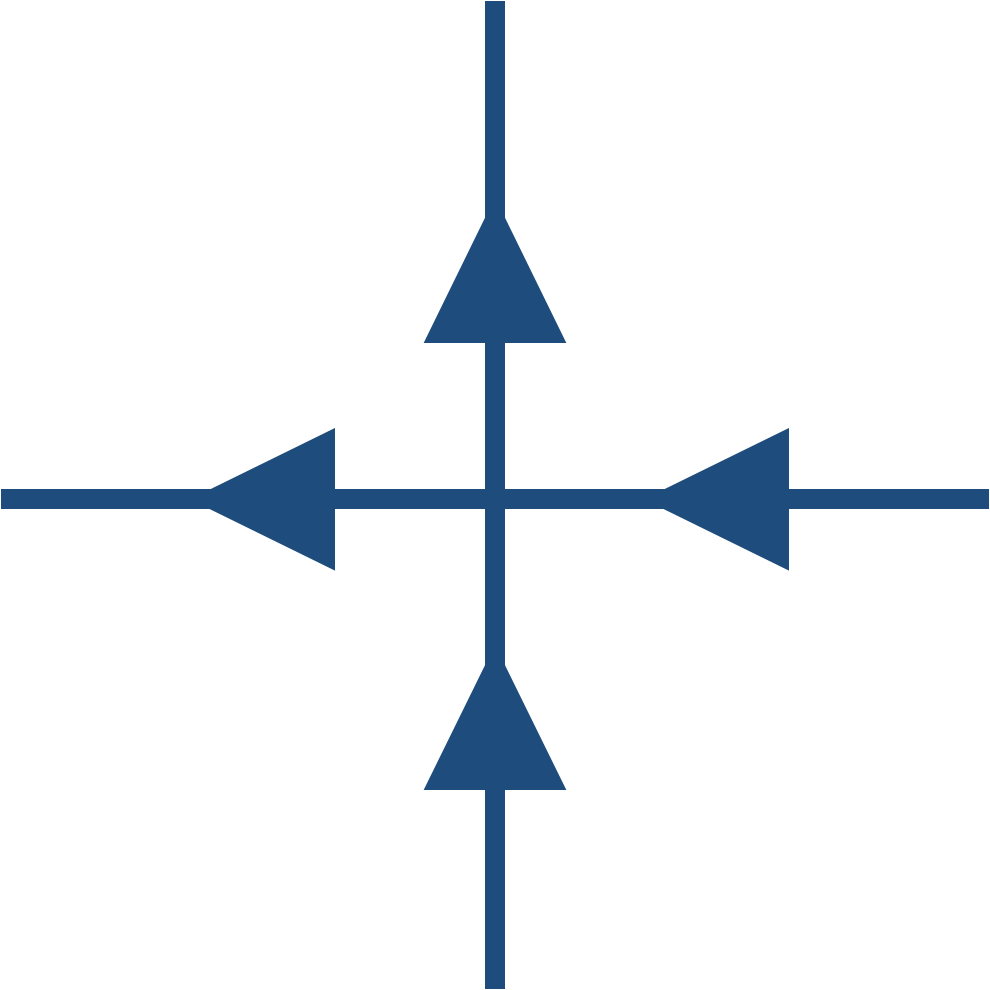} 
			& \includegraphics[width=0.07\textwidth]{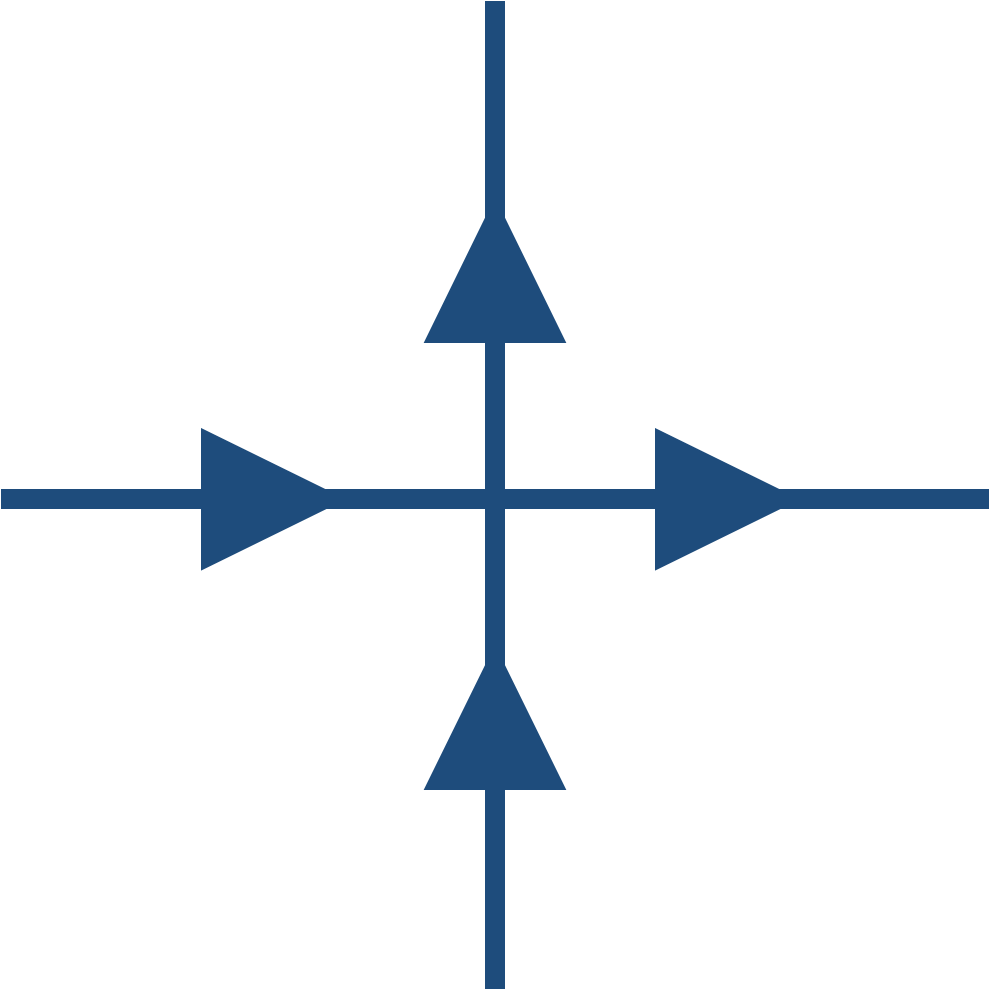} 
			& \includegraphics[width=0.07\textwidth]{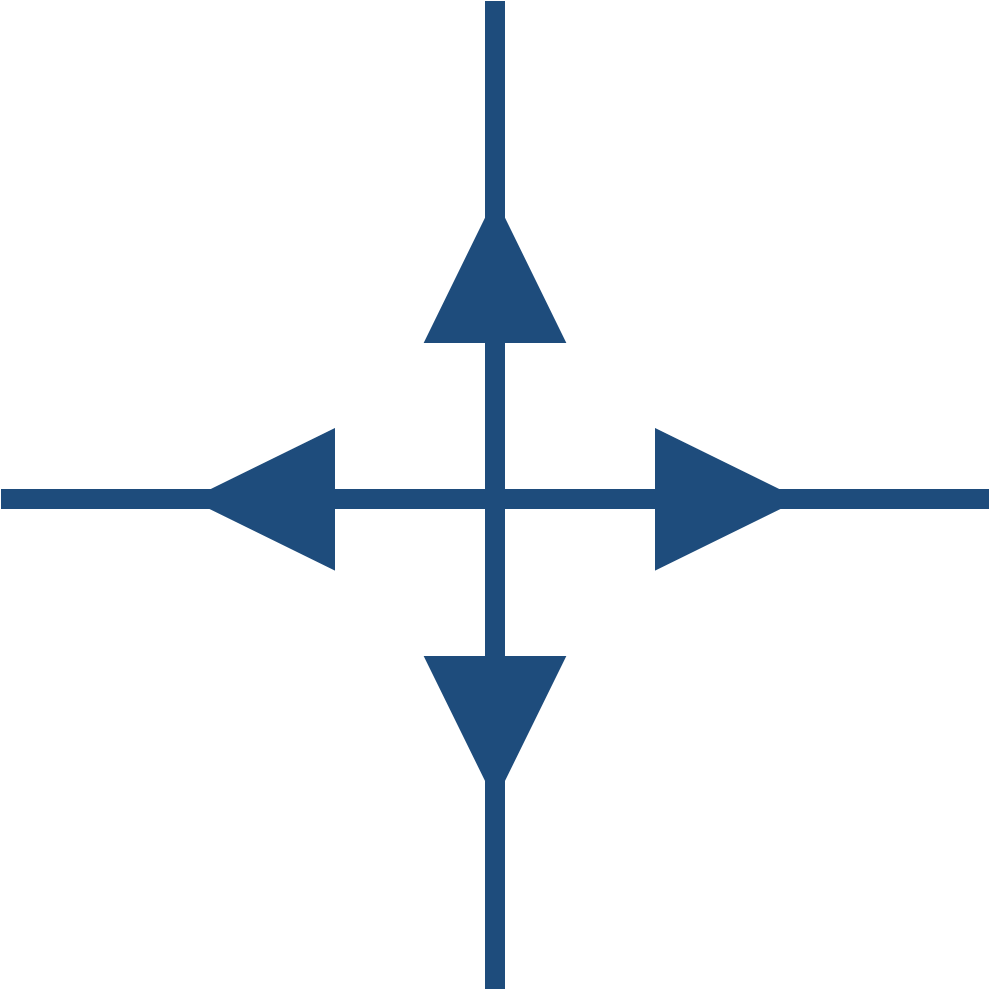} 
			& \includegraphics[width=0.07\textwidth]{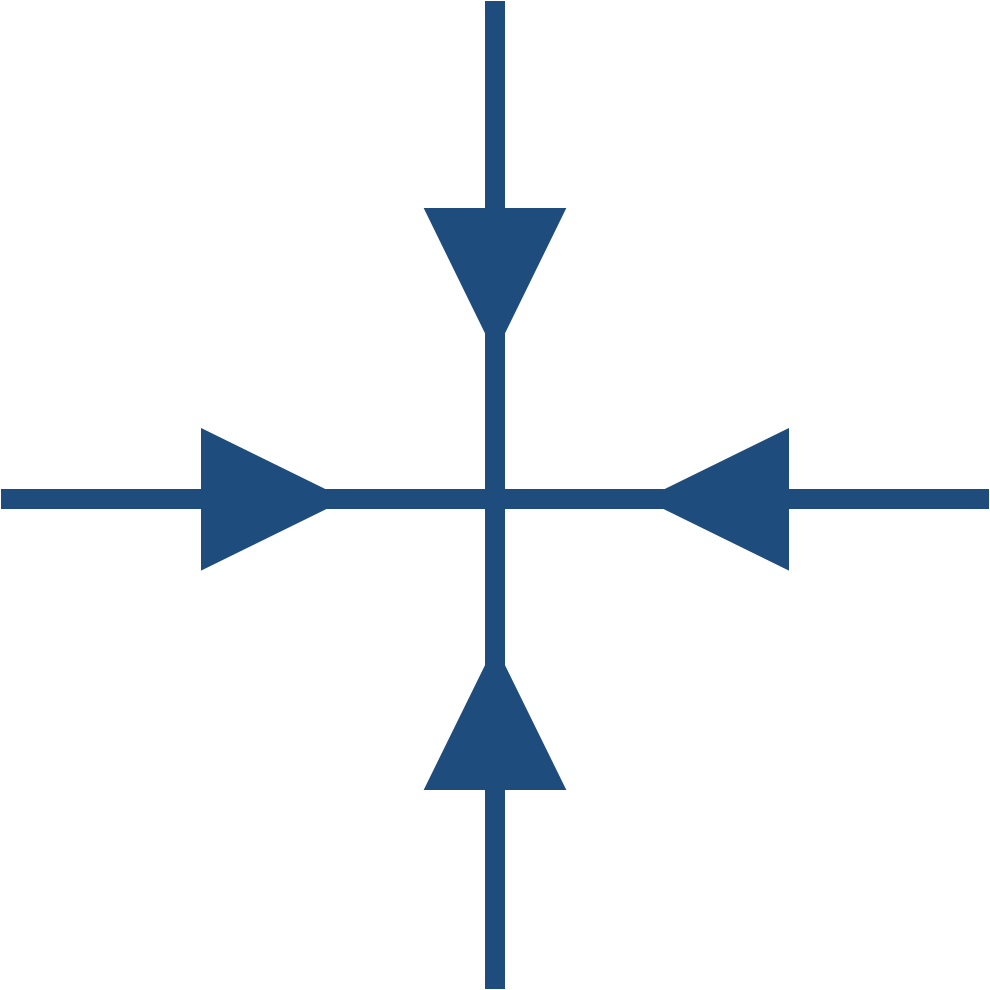} 
			& \includegraphics[width=0.07\textwidth]{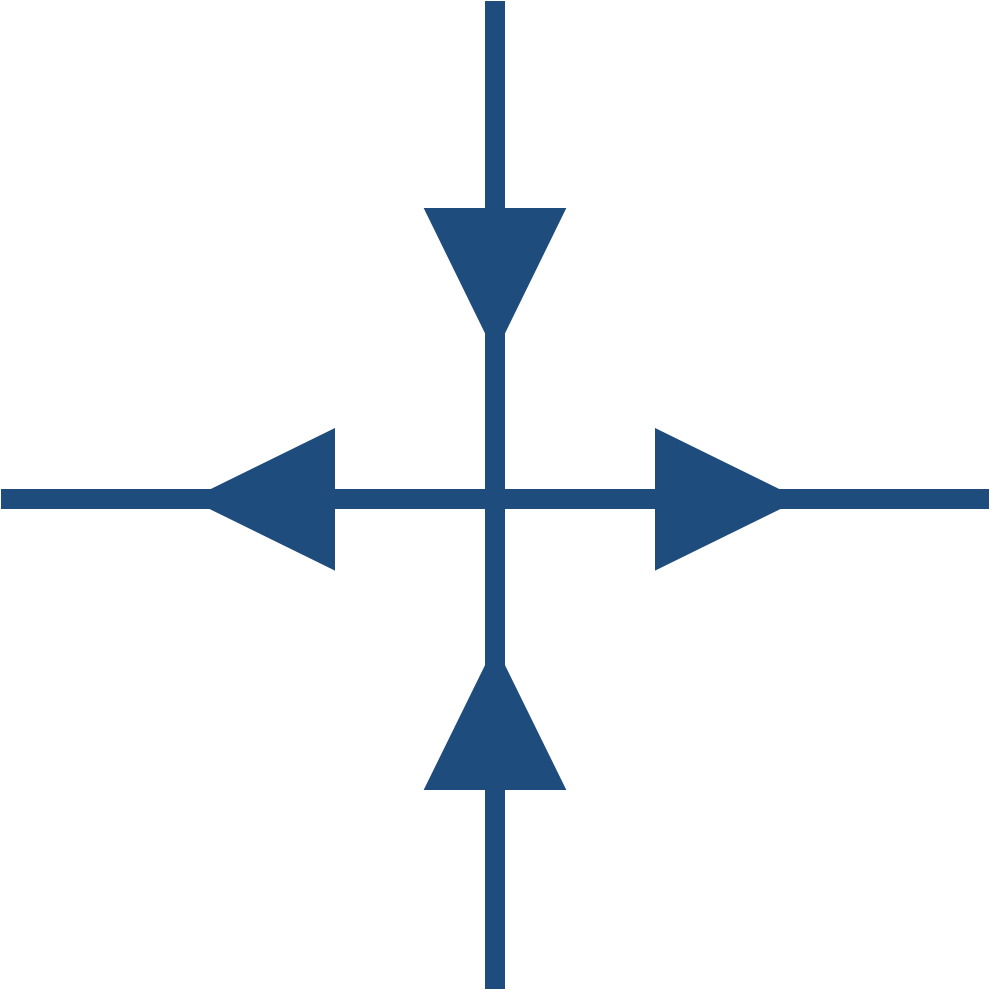} 
			& \includegraphics[width=0.07\textwidth]{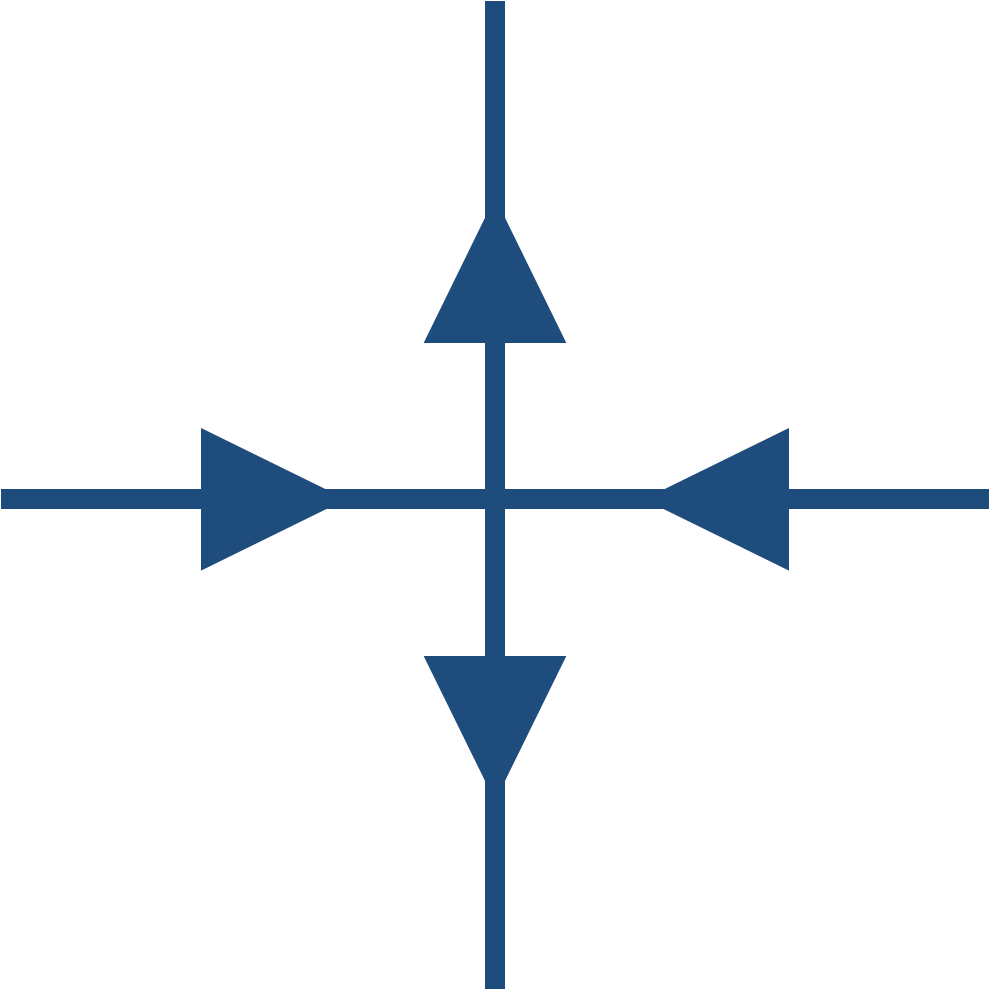} \\
			\hline\hline
			winding number $N_\mathsf{w}$& 0 & 0 & 0 & 0 & 1 & 1 & -1 & -1 \\
			\hline
			net flux $C_\mathsf{n}$& 0 & 0 & 0 & 0 & 4 & -4 & 0 & 0 \\
			\hline
			flux in $x$-direction $C_\mathsf{x}$& 0 & 0 & 0 & 0 &  2 & -2 & 2 & -2 \\
			\hline
			flux in $y$-direction $C_\mathsf{y}$& 0 & 0 & 0 & 0 & 2 & -2 & -2 & 2 \\
			\hline
		\end{tabular}
	\end{center}
	\caption{Vertex configurations in eight-vertex model and their winding numbers around the vertex center, total fluxes and fluxes in $x-$ and $y-$directions.} 
	\label{Fig_8vM}
\end{figure*}

The square-lattice eight-vertex model is a canonical 
 exactly solvable model	
\cite{Sutherland1970,FanWu1971PRB,Baxter1971,Kadanoff1971PRB,baxter2007exactly}.
It is constructed by placing a binary arrow (left/right or up/down) on every edge of the square lattice,
but only allowing vertex configurations of even number of arrows pointing in/out.
The eight allowed vertex configurations are shown in Fig.~\ref{Fig_8vM}. 
Under open boundary condition, 
each vertex can be independently assigned an energy cost $E_i\;(i=1, ..., 8)$
in the most generic case. 
Specifying $E_i$ completes the definition of the classical model.

The eight-vertex model can be reformulated as an equivalent 
spin model that involves up to four-spin interactions
\cite{baxter2007exactly}.
The classical fracton model (Eq.~\eqref{Eqn_Ham_Fracton}) described in Sec.~\ref{SEC_II_Fracton_model}
is a special case of the more general equivalence.
The prescription of the duality is given below.

\begin{figure}[ht]
	\centering
	\subfloat[\label{Fig_vertex_dual_1_index}]
	{\includegraphics[width=0.1\textwidth]{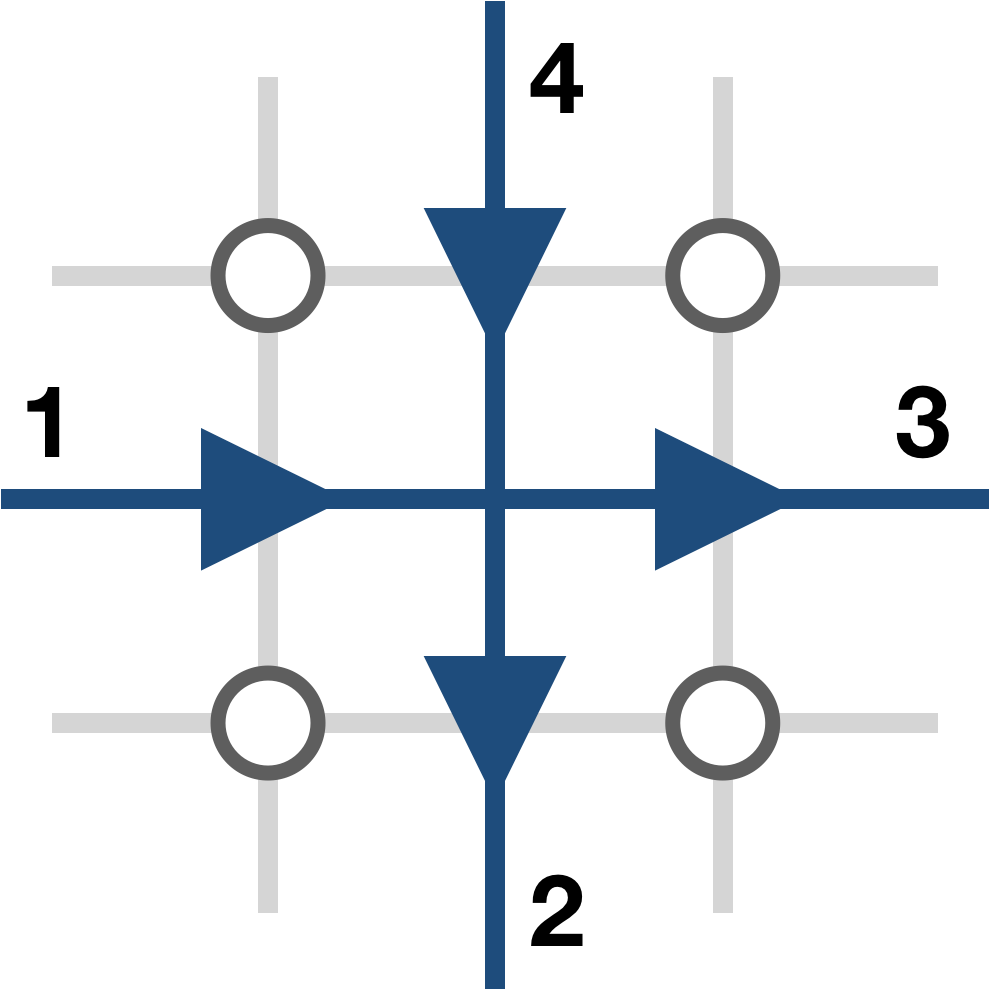}}\quad
	\subfloat[]{\includegraphics[width=0.1\textwidth]{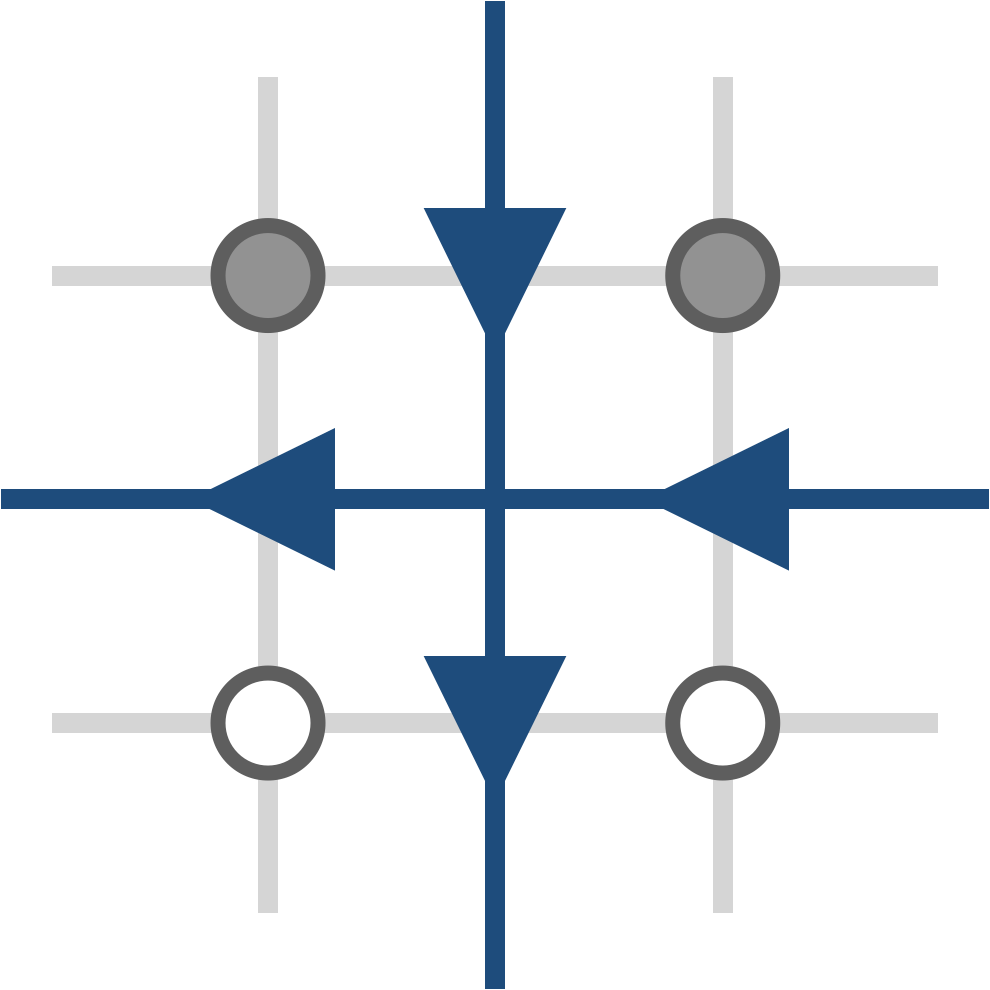}}\quad
	\subfloat[]{\includegraphics[width=0.1\textwidth]{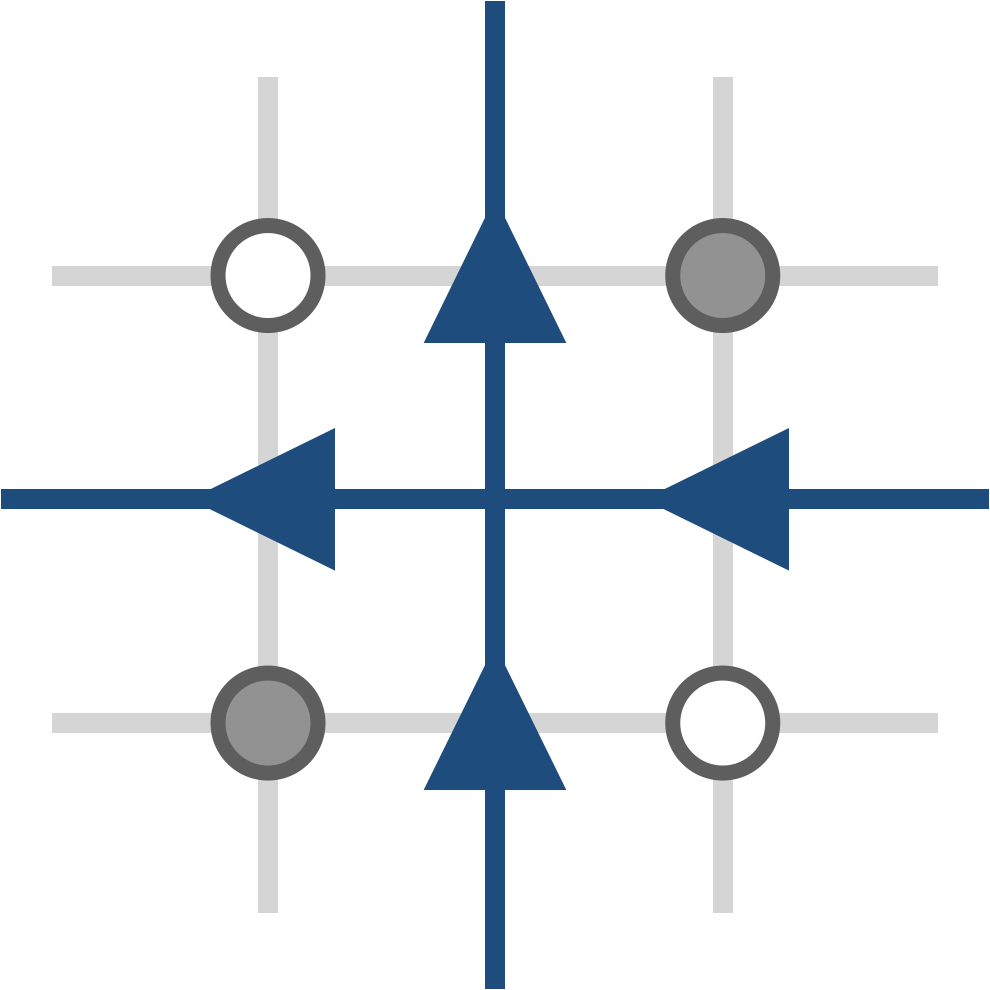}}\quad
	\subfloat[]{\includegraphics[width=0.1\textwidth]{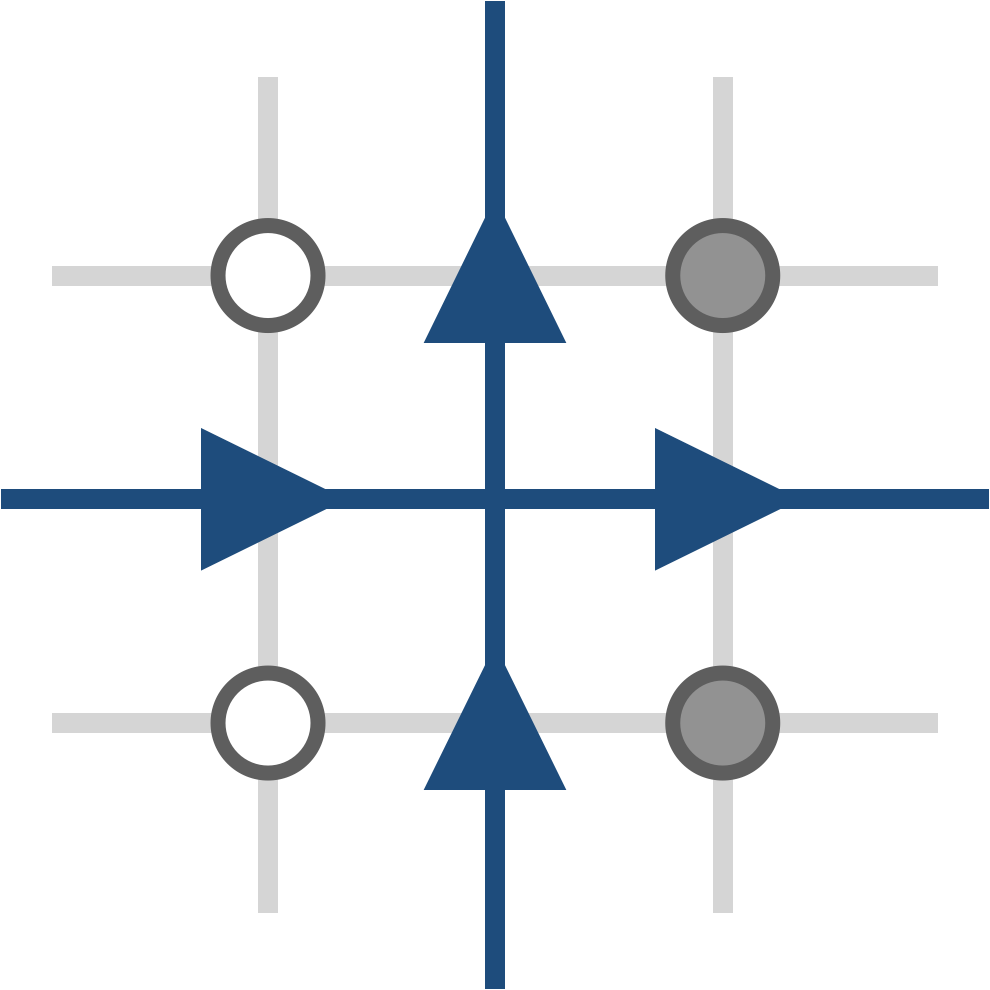}} \\
	\subfloat[]{\includegraphics[width=0.1\textwidth]{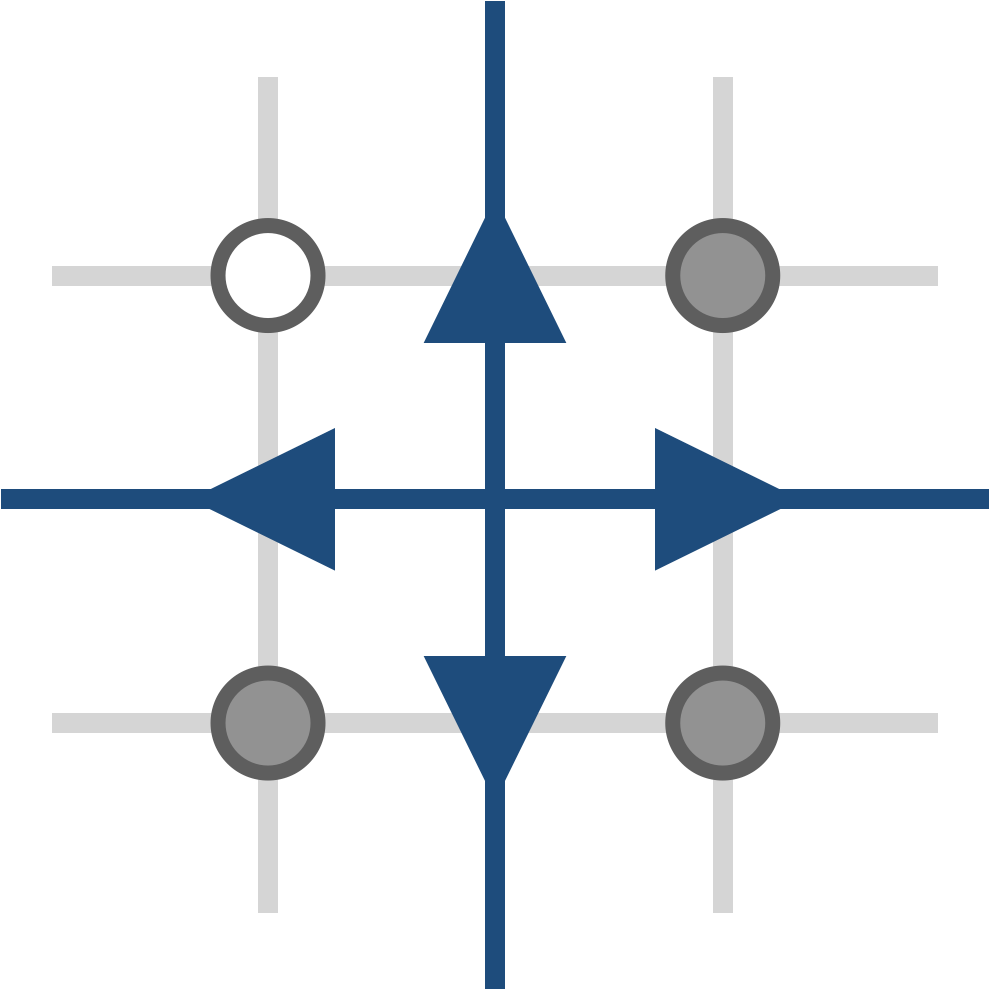}}\quad
	\subfloat[]{\includegraphics[width=0.1\textwidth]{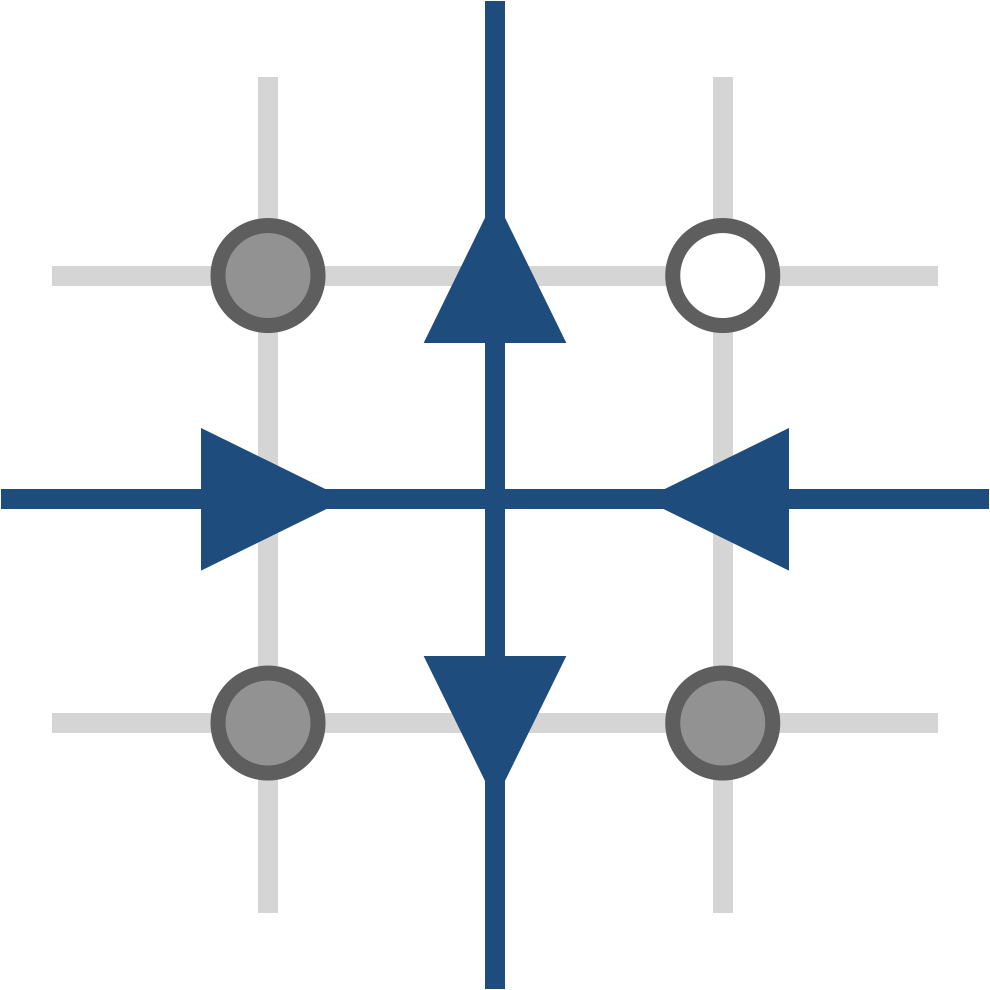}}\quad
	\subfloat[]{\includegraphics[width=0.1\textwidth]{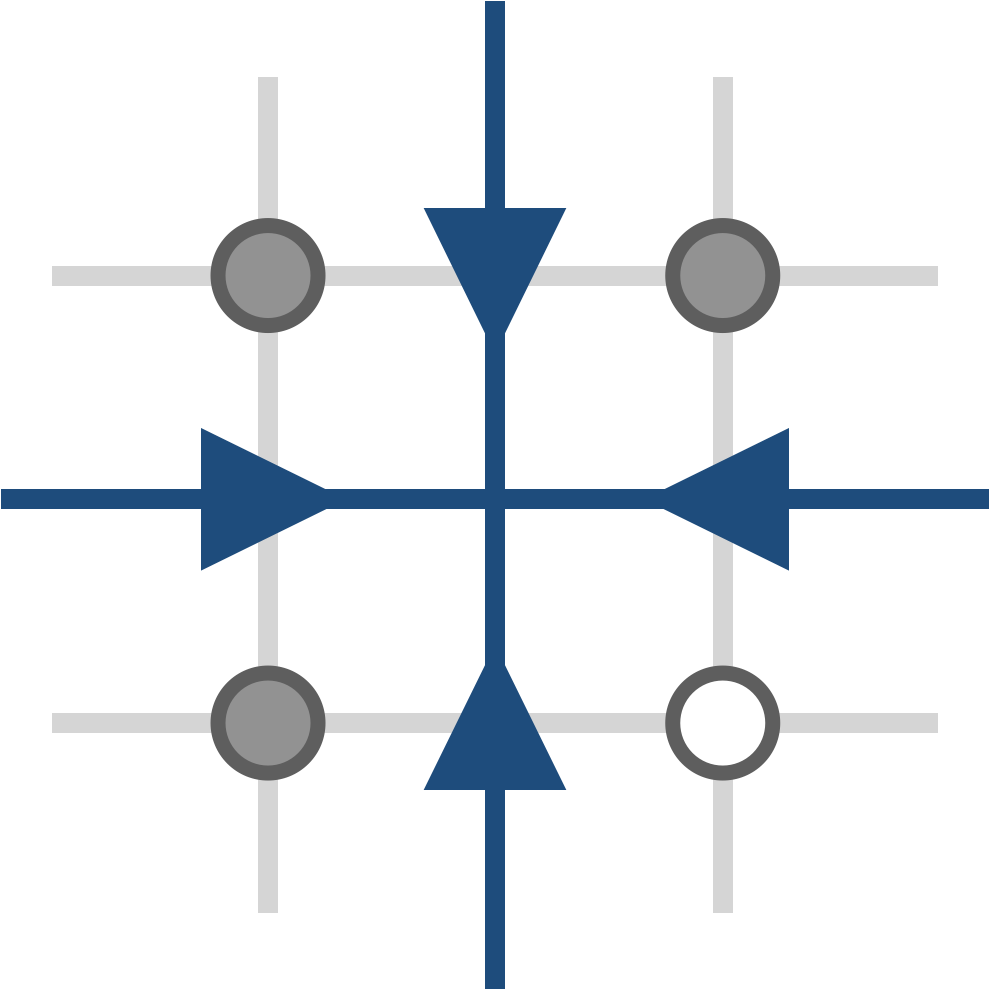}}\quad
	\subfloat[]{\includegraphics[width=0.1\textwidth]{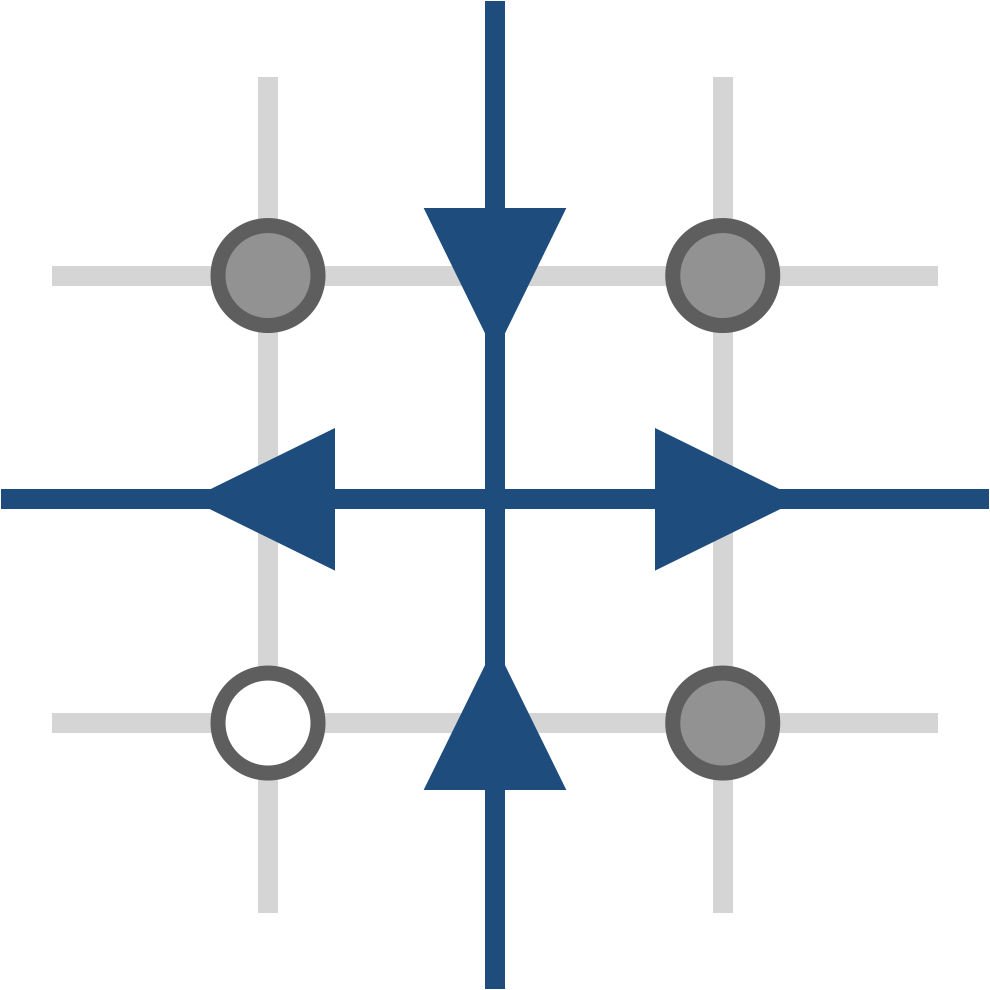}} 
	\caption{Mapping between the spin configurations and the vertices.
		(a-d): Spin configurations of ground states $\mathcal{O}_p=1$ correspond 
		to vertices with zero winding number. Their energy cost is $E_i=-1$.
		(e-h): Spin configurations of Fractons $\mathcal{O}_p=-1$ correspond 
		to vertices with winding number $\pm 1$. Their energy cost is $E_i=+1$.
		The correspondence is two-to-one, since 
		flipping all spins maps to the same vertex.
	} 
	\label{Fig_vertex_dual}
\end{figure}
The eight-vertex model is defined on
the dual square lattice of the original fracton model.
The mapping between the arrow and spin configurations 
is illustrated in Fig.~\ref{Fig_vertex_dual}. 
Each edge of the dual lattice 
neighbors two spins of the original lattice, at the ends of
the perpendicularly intersecting edge. 
%
The arrow of the dual edge points right or down 
if the two spins are aligned in the same direction,
and left or up otherwise.
Such assignment guarantees that
any four-spin configuration is mapped to one of the
eight vertices listed in Fig.~\ref{Fig_8vM}.
The mapping has a global two-fold degeneracy: the vertices remain the same after flipping all spins.

The dual Hamiltonian for the eight vertex model is
\begin{equation}
		\mathcal{H}_\textsf{EV} = - \frac{1}{2} \sum_v (\sigma_1 \sigma_3 + \sigma_2 \sigma_4)			.
\end{equation}
Here $v$ denotes all vertices in the dual lattice,
and $\sigma_i$ is the value of arrows on edge $i$,
defined as
\begin{equation}
\sigma_i = 
\begin{cases}
1 \qquad & \text{if it points right or down;}  \\
-1 \qquad & \text{if it points left or up.} 
\end{cases}
\end{equation}
The assignment of subscripts $1,2,3,4$ around a vertex is shown in Fig.~\ref{Fig_vertex_dual_1_index}.
Note that this is not equivalent to a bunch of non-interacting 1-dimensional spin chains,
since the constraint of eight-vertex configuration is enforced.

The vertices of winding number zero (cf. Fig.\ref{Fig_8vM}) correspond to the 
ground-state spin configurations of 
\begin{equation}
\mathcal{O}_p=1,
\end{equation}
and have energy cost 
\begin{equation}
E_i = -1\;, i = 1,2,3,4.
\end{equation}
Those of winding number $\pm 1$
correspond to the spin configurations of 
\begin{equation}
\mathcal{O}_p=-1,
\end{equation}
and have energy cost 
\begin{equation}
E_i = +1\;, i = 5,6,7,8  \;,
\end{equation}
which agree with the original fracton model (Eq.~\eqref{Eqn_Ham_Fracton}).
The prescription of the duality is concluded here.\\


The dual eight-vertex model has the advantage
of illustrating various concepts of fracton models.

\begin{figure}[ht]
	\centering
	\subfloat[]{\includegraphics[width=0.2\textwidth]{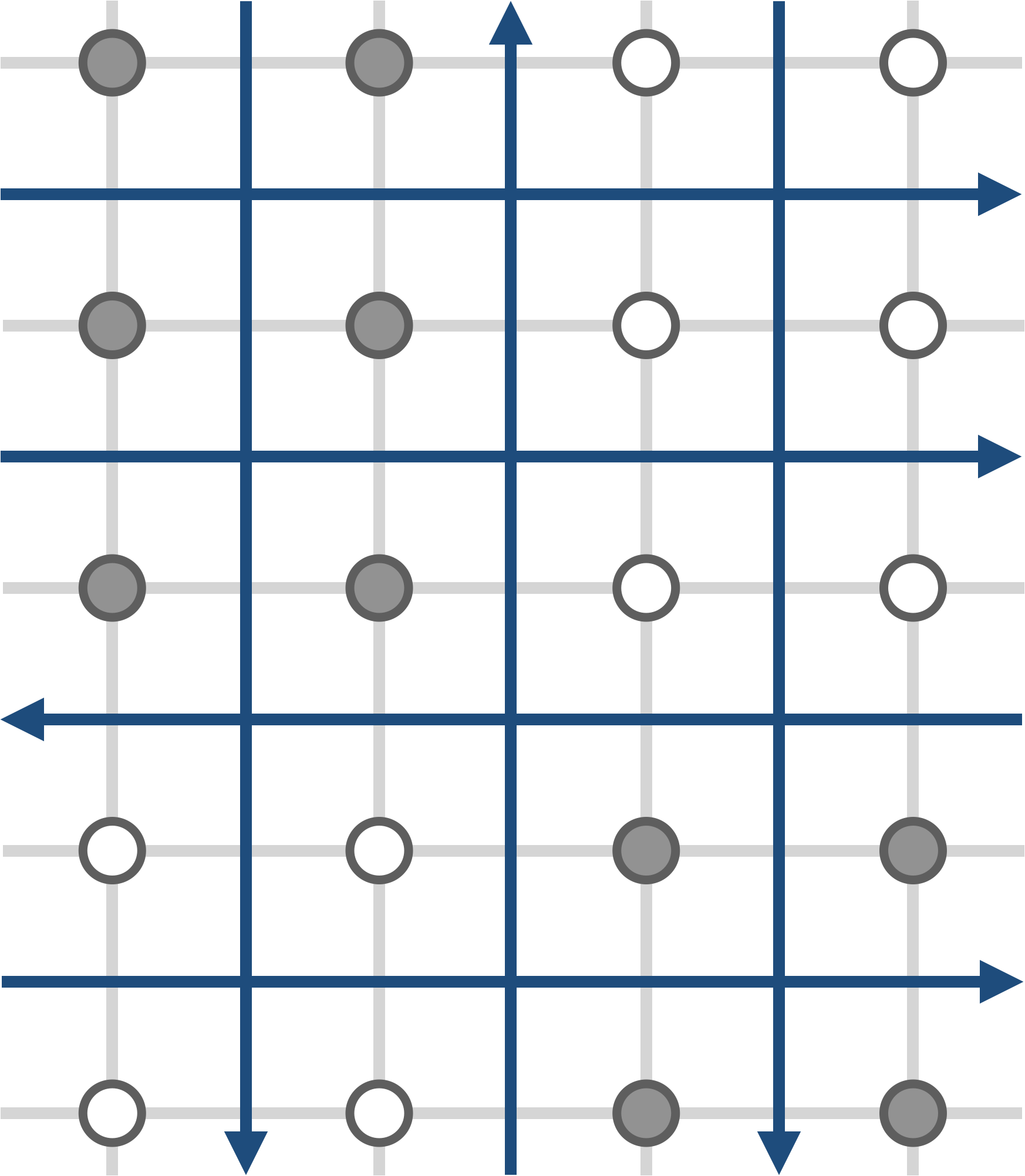}}\quad
	\subfloat[]{\includegraphics[width=0.2\textwidth]{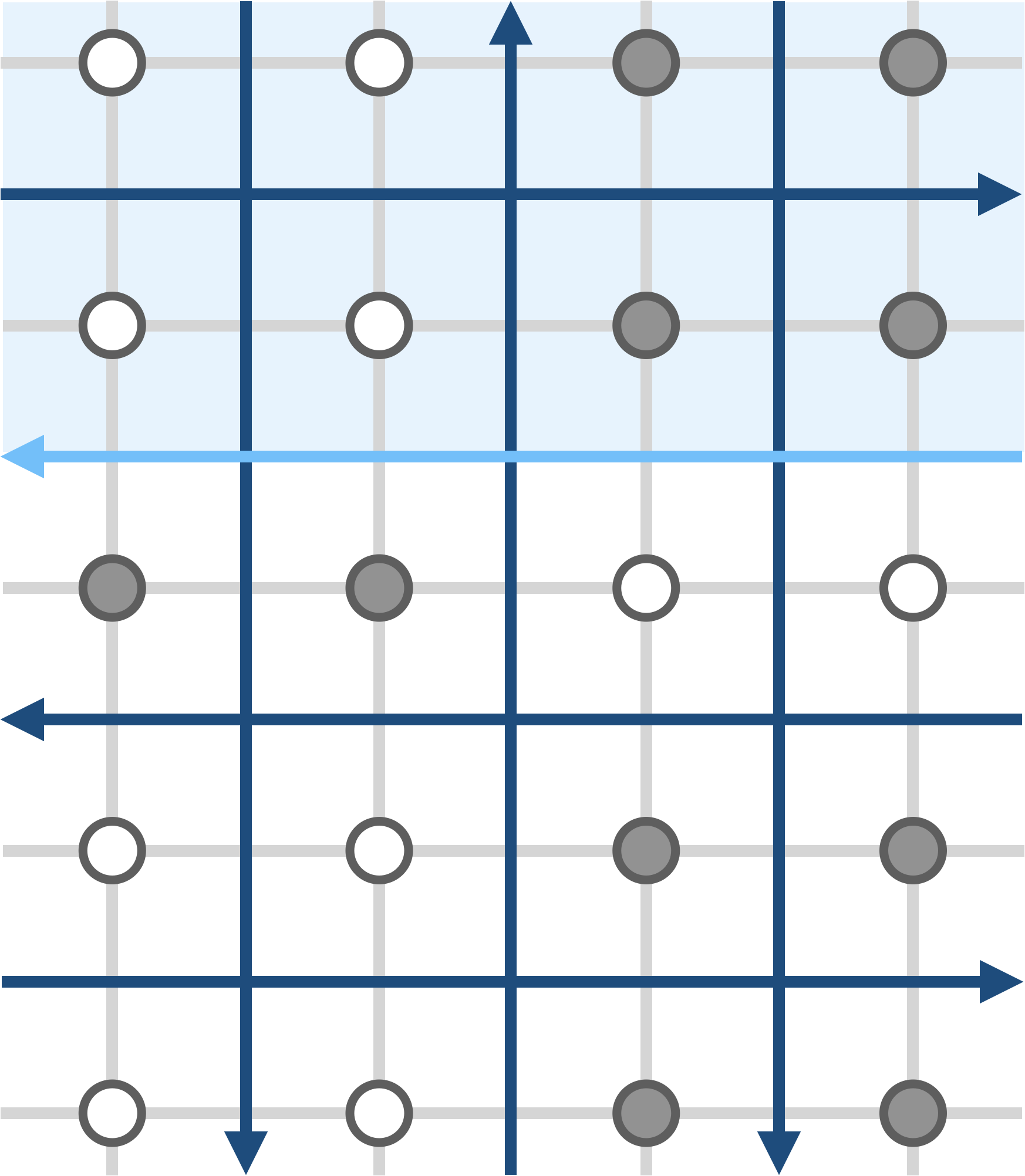}}
	\caption{Ground state degeneracy in the two dual models.
		(a): A spin configuration of ground state, and its dual eight-vertex model
		state. 
		In the eight-vertex model the ground state is such that all arrows on the same
		line align in the same direction.
		(b): Another spin configuration of ground state, obtained from (a) by flipping all spins
		in the light-blue shaded area.
		In the dual eight-vertex model it corresponds to flipping a line of arrows.
	} 
	\label{Fig_lat_vertex_dual}
\end{figure}
Firstly let us examine the ground state degeneracies.
In the dual eight-vertex model, the ground states become very simple:
all arrows on the same straight line have to align in the same direction.
The action of flipping all spins on one side of a straight line
corresponds to flipping  the arrows of the entire line.
This is illustrated in Fig.~\ref{Fig_lat_vertex_dual}.
The ground state ensemble is thus equivalent to 
a number of uncorrelated Ising spins on the boundary,
which makes it apparent that its entropy is proportional to
the  boundary area.

Next we turn to the fracton excitations. 
The dual model illuminates 
a qualitative difference 
between the its effective theory --- rank-two U(1) gauge theory (the traceless, scalar charged version) ---
and conventional U(1) gauge theory in two-dimensional space.
The rank-two U(1) gauge theory
accounts for the topological excitations of non-zero \textit{winding number} $N_\mathsf{w}$
of the underlying vector field,
while the conventional $U(1)$ gauge theory 
accounts for the non-zero net flux  $C_\mathsf{n}$.

As one can see in Fig.~\ref{Fig_vertex_dual},
the fractons (vertices $5,6,7,8$ in Fig.~\ref{Fig_8vM}), or ``charge'' of the rank-two electric field,
are actually  vertices with winding number $\pm 1$.
In contrast, in the conventional electromagnetism,
the ``charge'' is the net flux of the underlying electric field,
or just the charge as we know it (vertices $5,6$).

The observation echoes the fracton-elasticity duality \cite{Pretko2018PRL},
where the underlying vector field is the lattice distortion,
and disclinations corresponds to a non-zero winding of the distortion \cite{Beekman2017}.

The dual eight-vertex model is also an elegant demonstration 
of the subsystem symmetries and charges discussed in \cite{Vijay2016,Shirly2019SciPost}.
Each Fracton vertex will introduce an $x-$ and $y-$ subsystem 
charge $C_\mathsf{x}$ and $C_\mathsf{y}$ on the $x-$ and $y-$direction line it is located.
The charges are the flux in $x$ and $y$ listed in Fig.~\ref{Fig_8vM}.
They are related to the winding number by
\begin{equation}
N_\mathsf{w} = \frac{C_\mathsf{x}C_\mathsf{y}}{4}			.
\end{equation}

Two different lines
have their independent charges.
The total charge of each line, which can be $0$ or $\pm 2$,
must be conserved by local spin flipping.
Therefore a single fracton is completely localized,
since moving it will change the subsystem charges.
A two-fracton bound state can move in
$x-$direction if they give zero charge on the $y-$direction lines.
A four-fracton bound state has zero subsystem charge on any line,
hence is free to move.

\section{Hyperbolic Dual Eight-Vertex Model } \label{SEC_4_Vertex_model_hyperbolic}

\begin{figure}[ht]
	\centering
	\includegraphics[width=0.4\textwidth]{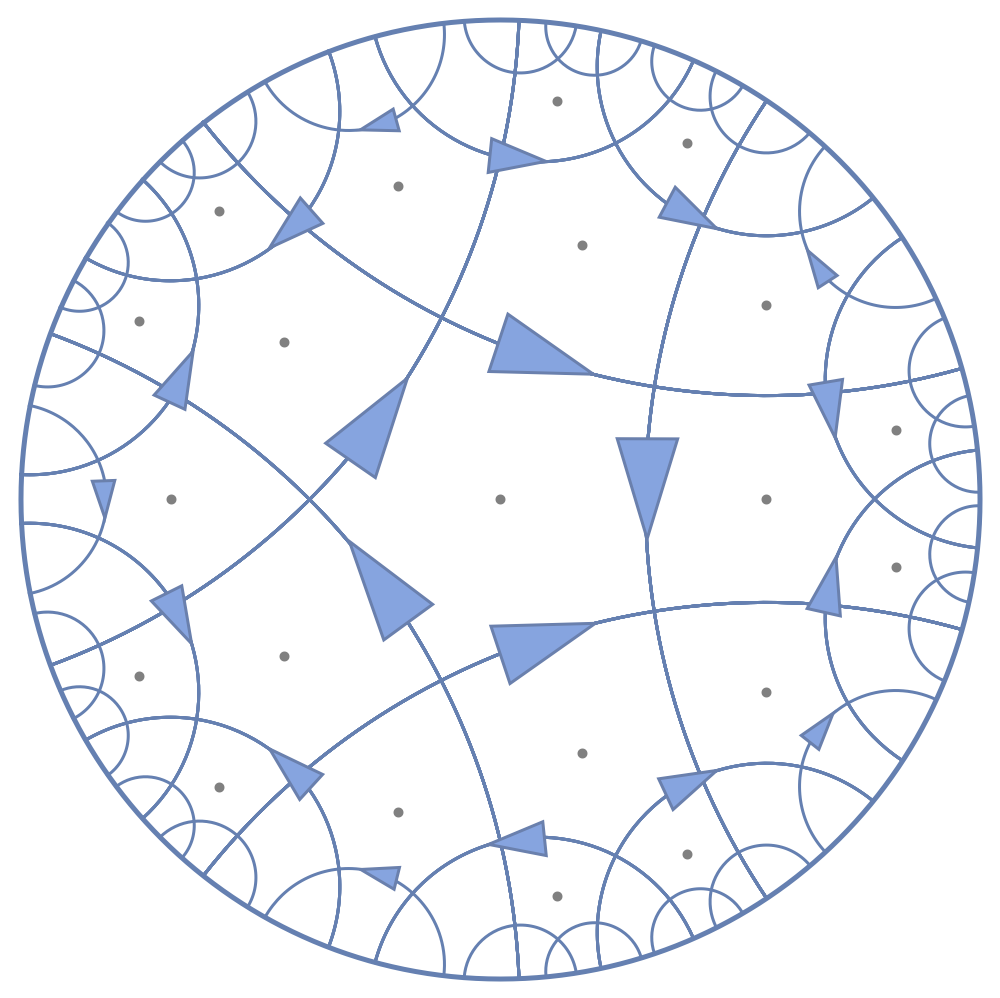}
	\caption{Dual eight-vertex model on the hyperbolic disk at $T=0$.
		Each geodesic carries an independent binary arrow.
	} 
	\label{Fig_Bit_Thread}
\end{figure}

The eight-vertex model dual to the hyperbolic fracton model 
is obtained by simply upgrading the square lattice  to 
the $(5,4)$ tessellation of the hyperbolic disk.
In the dual model, each pentagon's  edge has an associated binary arrow,
and vertices are still restricted to
the eight configurations in Fig.~\ref{Fig_8vM}.
Here we assign the arrow directions in the following way:
We start from the obvious fracton model ground state of  
all spins pointing up.
We then define the corresponding vertex model
configuration is that (1) all arrows on the same geodesic align 
in the same direction; (2) the arrow on the geodesic flows clock-wise.
All other vertex states are fixed following these rules.

For the ground state, all edges on the same geodesic have aligned
arrows. 
Flipping all spins on one side of a geodesic
corresponding to flipping its arrow direction.
Fig.~\ref{Fig_Bit_Thread} shows one example of ground state eight-vertex model configurations.
For fracton excitations,
the concept of subsystem charges for each geodesic is also still valid.

\section{Bit-Thread Realization}
\label{Sec_5_Bit_thread}
When restricted to its ground states, 
the dual eight-vertex model becomes
a collection of geodesics, each associated with a binary arrow.
This is a simple discrete and classical realization of the 
\textit{bit-thread model} proposed in Ref.~\cite{Freedman2017CMaPh}
as a powerful conceptual tool to visualize holography.

In the bit-thread model, 
the elementary physical object 
is a divergence-free vector field
in the bulk with pointwise bounded norm,
referred to as the \textit{flow}.
Like how physicists visualize electric/magnetic fields,
the flow lines can be viewed as threads.
Each thread carries an independent bit of information (or 
two entangled qubits),
and stretches from one boundary point to another.
The full-fledged geometric theory of the bit-thread model 
is able to account for various properties of holographic entanglement entropy.
For example, 
since the covering geodesic of boundary subregion $A$ 
is the narrowest bottleneck separating $A$ and its complement $A^c$,
it sets the upper bound of the entanglement entropy between them.
Following the max-flow min-cut  principle \cite{Freedman2017CMaPh},
this upper bound is saturated,
so that 
the entanglement entropy obeys the RT formula.

In the  eight-vertex model at zero temperature,
each   geodesic is a thread or discretized flow,
and carries the binary arrow as 
one bit of classical information.
The bit threads visualize
the mutual information between two subregions.
It is simply counted by
how many geodesics the two subregions
share,
as both subsystems can measure the directions of these arrows.

\begin{figure}[ht]
	\centering
\includegraphics[width=0.35\textwidth]{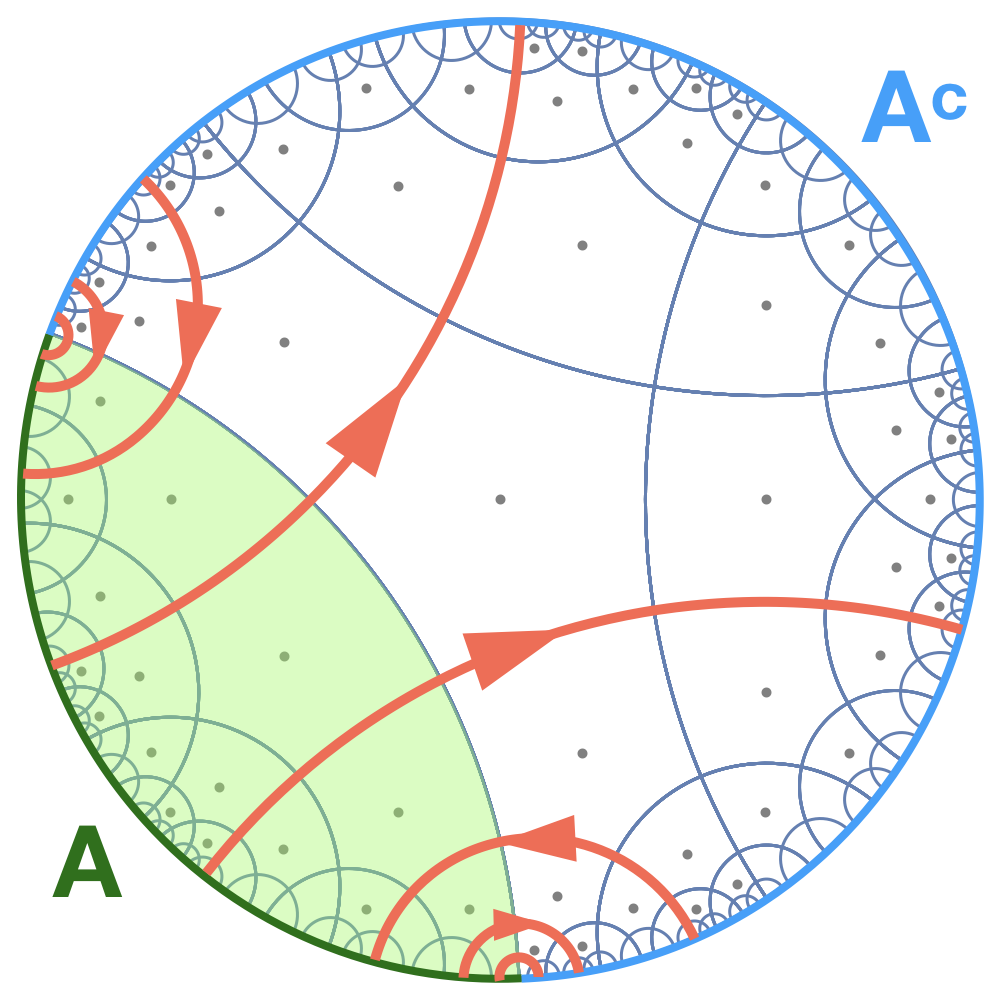}
	\caption{Bit-Thread Realization and Ryu-Takayanagi formula
		for mutual information.
		Each geodesic is a thread carrying a bit of information.
		The threads shared between two region (blue and green)
		saturates the bottleneck between them, i.e., 
		the covering minimal surface. 
		Hence the  mutual information obeys the RT formula.
	} 
	\label{Fig_RT_MI}
\end{figure}

The idea of the minimal covering surface being the bottleneck
is clearly represented in the eight-vertex model.
As shown in Fig.~\ref{Fig_RT_MI}, 
the   geodesics highlighted in orange
are the threads carrying the mutual information from boundary segment $A$ (green) to its complement $A^c$ (blue). 
It is straightforward to identify that the minimal covering surface, or the geodesic homologous to $A$, is the bottle neck of the orange region-crossing threads,
which is exactly the picture described in the bit thread model.

The bit-thread model realization is simple, 
yet bears some non-trivial implications.
We know that the rank-2 U(1) theories are linearized limit of
certain gravitational theory,
and the toy fracton model here is a discretized and Higgsed 
version of the rank-2 U(1) theories.
By studying the field theory and utilizing the duality
established here, 
it might be possible to derive the full bit-thread model
from (linearized) gravity.
This would be an interesting result
for  holographers.

We noticed a recent development yields very similar results. 
In Ref.\cite{Jahn2019arXiv}, Jahn etc. studied the holographic 
tensor network in the language of majorana dimers,
and discovered that the tensor networks have the same picture
as we described here --- entangled EPR pairs are 
linked by bit threads that form the hyperbolic lattice.
This is a very strong indication of hidden connections between
fracton models and holographic tensor networks.

\section{Bulk-Boundary Isometry for Diluted Fracton Excitations}
\label{Sec_6_Isometry}
Isometry is a core issue for toy models of holography \cite{Pastawski2015,Yang2016,Hayden2016,Qi2018}.
In the context of the classical fracton toy models,
roughly speaking, isometry means to require that the boundary can
unambiguously determine the bulk.
It can be rigorously defined as :

\textbf{Definition: }
A subset of all possible spin/vertex states 
is \textit{isometric}, if none of its two elements have the same boundary
state.

That is to say, within the chosen subset 
of all possible spin/vertex states, \
the boundary state uniquely determines the bulk.
Of course, the subset has to be a sensible choice --
normally we would expect it to contain many low-energy
states.
For example, if it is the set of all the ground states,
then isometry holds exactly.

\begin{figure}[t]
	\centering
	\includegraphics[width=0.45\textwidth]{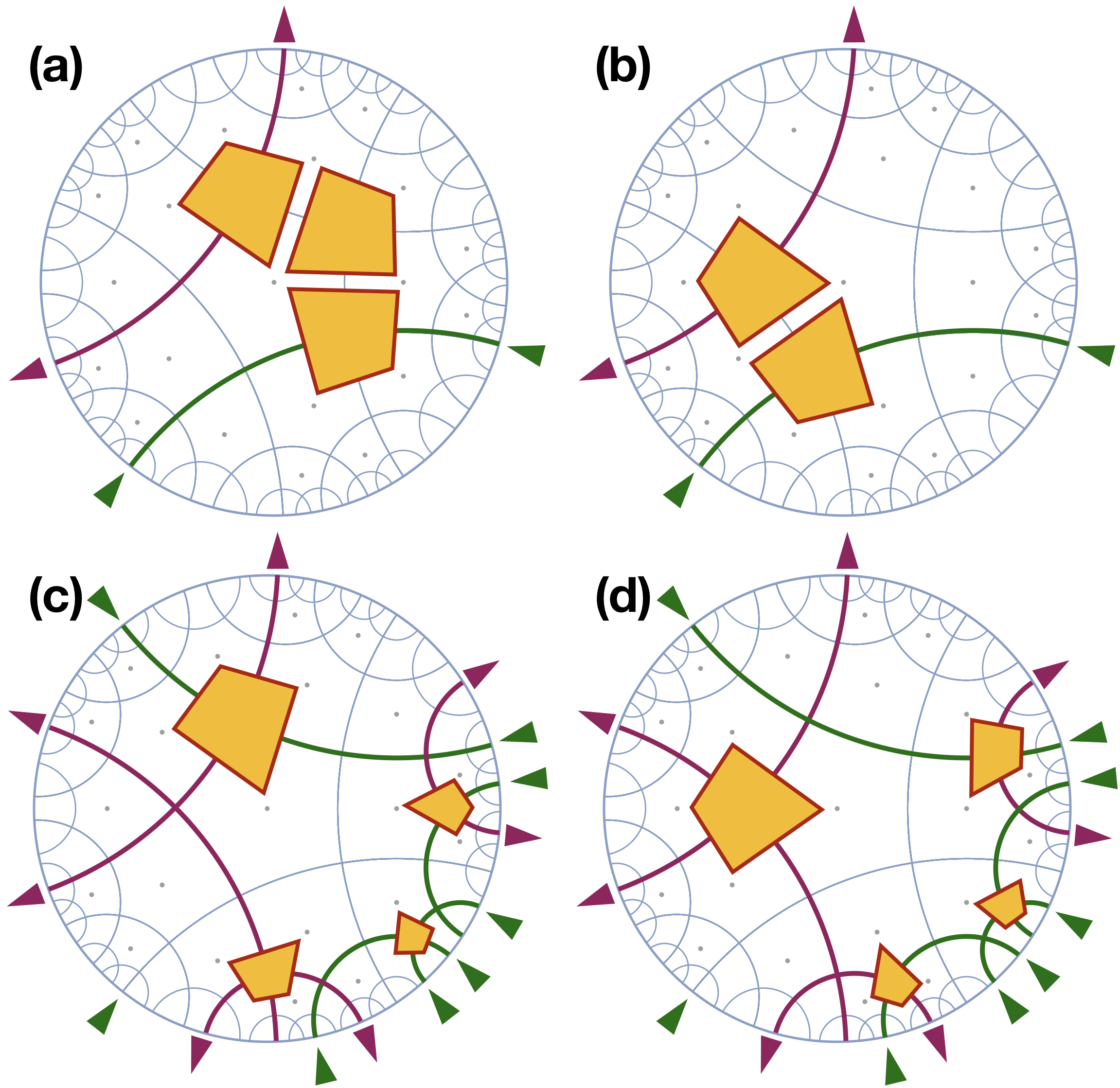}
	\caption{Two examples of isometry violation.
		(a, b) The dense three-fracton excited states cannot be distinguished
		from the two-fracton excited states from the boundary.
		(c, d) Two states with four-fracton excitations cannot be distinguished from the boundary.
	} 
	\label{Fig_Iso_violation}
\end{figure}

If the subset includes certain 
configurations at higher energies,
the isometry will eventually break down.
Two examples are given in Fig.~\ref{Fig_Iso_violation}.
This means the violation of holography,
but is acceptable.
Because for toy models, it is often
the case that isometry 
(and thus holography) only holds at low energy. 
After all, the AdS geometry will be distorted beyond small perturbations
by local high energy excitations,
which is not captured by the toy models at all.

The question now becomes:
how can we include more configurations
at higher energy levels
but maintain isometry?
Or equivalently,
if we include all states below a certain energy level,
how much is the isometry broken?

\begin{figure}[ht]
	\centering
	\includegraphics[width=0.45\textwidth]{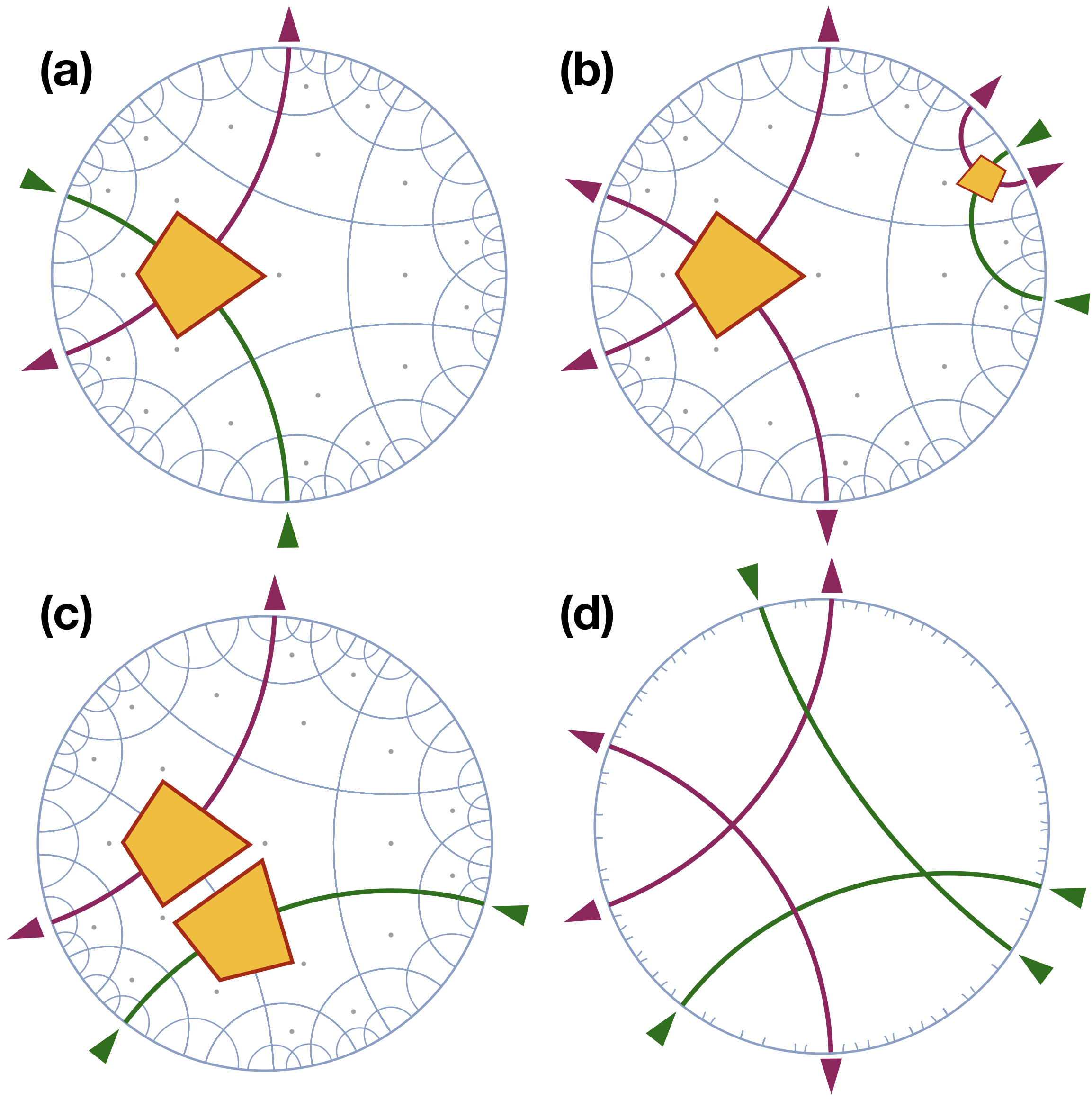}
	\caption{
		(a) A single fracton excitation can be reconstructed from the boundary
		by identifying the geodesics with non-zero subsystem charges.
		(b, c) Two-fracton excitations can also be reconstructed, even if 
		they lie on the same geodesic which has zero-charge from boundary point of view. Because such geodesic can be uniquely identified.
		(d) Geodesics in this configuration are forbidden by the lattice geometry,
		which guarantees situations of (b) are always unambiguous about 
		the locations of two fractons.
	} 
	\label{Fig_Isometry}
\end{figure}

To start with,
including all single fracton excited states does not break isometry.
This is almost obvious,
but we still analyze it in the eight-vertex picture
to pave way for more complicated situations.
As we discussed in Sec.~\ref{SEC_3_Vertex_model_square},
each geodesic has its own subsystem charge.
A fracton will introduce non-zero subsystem charges 
to the two geodesics $\gamma_1$ and $\gamma_2$ 
it sits on.
So the subsystem charge 
is zero if there are zero or even number of fractons
sitting on it, 
and $\pm 2$ if there are odd number of fractons
sitting on it.
In the case of a single fracton excitation,
by examining the boundary arrows,
we can identify $\gamma_1$ and $\gamma_2$
with $\pm 2$ charges,
thus determine the location of the fracton,
and the entire bulk  (Fig.~\ref{Fig_Isometry}a).

All the two fracton excited states
can also be included in this subset.
The most-likely case 
is that we have four geodesics
with non-zero subsystem charges,
which pins down the two fractons (Fig.~\ref{Fig_Isometry}b).
The hyperbolic lattice geometry guarantees
us that situation like Fig.~\ref{Fig_Isometry}d will never happen,
since in that case the four geodesics form
a rectangle with all its corners of angle $\pi/2$.
Such rectangles cannot exist in hyperbolic geometry.

The other possibility is 
when the two fractons sit on the same 
geodesic, a situation illustrated in Fig.~\ref{Fig_Isometry}c .
In this case there are only two geodesics
with non-zero subsystem charges.
However, due to the lattice geometry,
there is one and only one geodesic that intersects both,
so it can be uniquely determined.
Hence the two fractons' positions can always be located.

The isometry will be broken if we further
include all three-fracton excited states.
Figs.~\ref{Fig_Iso_violation}a,b illustrate one of such examples,
in which the three-fracton excited state 
has the same boundary as the two-fracton excited state.
This can be fixed by excluding the cases when
the three-fracton excitations are \textit{dense},
that is, they locate around the same pentagon.
Once such cases are removed from the subset,
so that only the diluted three-fracton excitations 
are included,
isometry is recovered.

The same procedure can be applied as 
higher-energy states are included:
if by local operations a state can be
turned into a lower energy one (Figs.~\ref{Fig_Iso_violation}a,b)
or one at the same energy level (Figs.~\ref{Fig_Iso_violation}c,d),
it should excluded in the subset.
In this way we include as many lower energy states possible
while maintaining isometry.
To enumerate all cases 
is a slightly tedious task,
but in principle achievable. 
Roughly speaking,
as long as the fracton excitations 
are ``diluted'',
isometry holds.
This is actually very sensible,
since high energy density
means distortion of the local space geometry,
where the lattice model is not a good representation anymore.

Coming back to the question in the beginning 
of this section,
at low energy levels, we can include 
most of the states without violating isometry.
Or, if we include all states at low energy levels,
the isometry is not broken too much.
This is also the case of  holographic tensor networks \cite{Yang2016}

An interesting side note
is that the mostly-preserved isometry 
for low energy excitations 
is a consequence of the negatively-curvatured geometry.
On the Euclidean lattice, 
isometry is completely violated starting from two fracton excitations.
\section{Non-Local Black Hole Microstate Degree of Freedom}
\label{Sec_7_BH}
\begin{figure}[ht]
	\centering
	\includegraphics[width=0.4\textwidth]{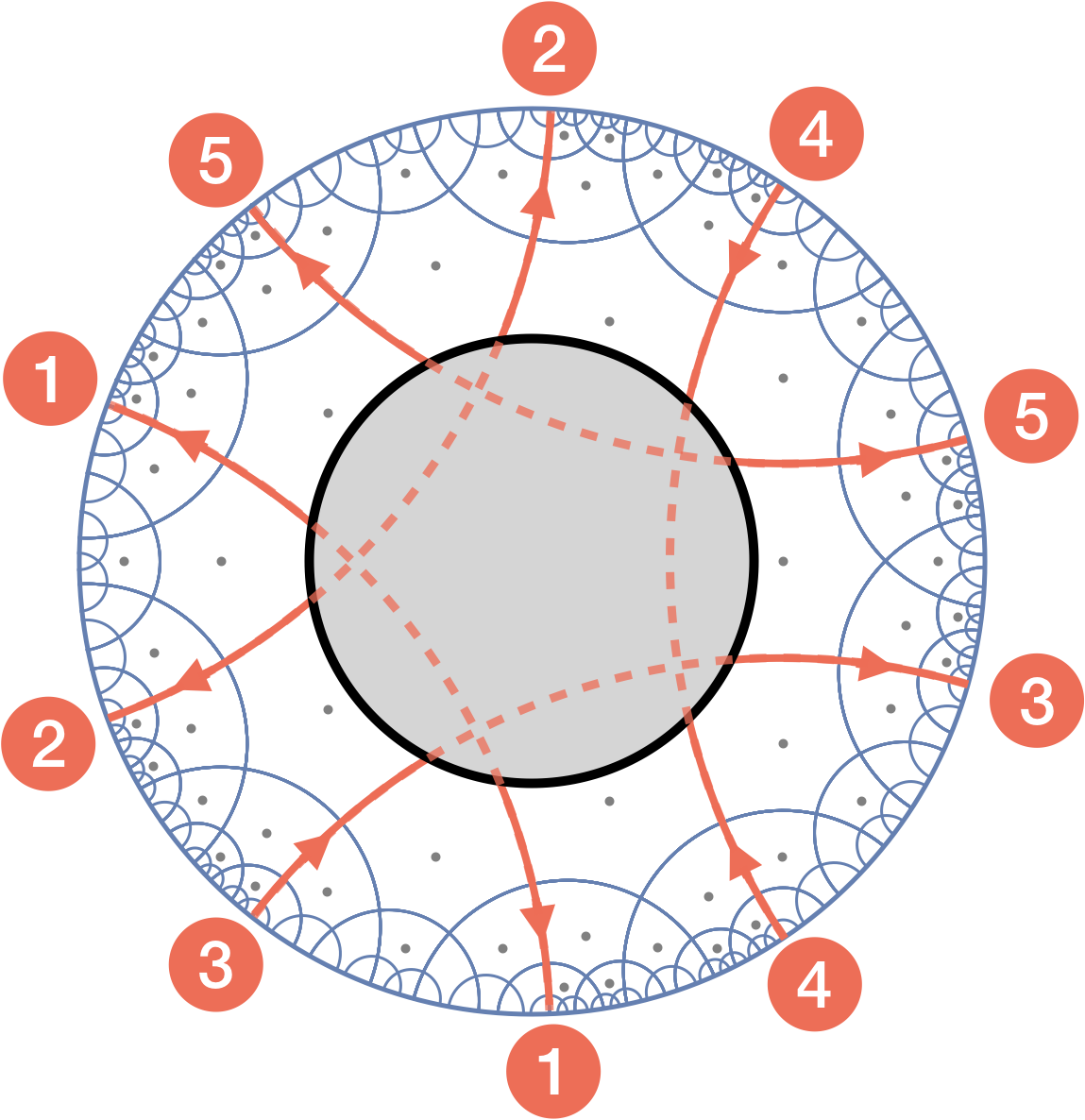}
	\caption{A black hole in the hyperbolic fracton model. 
		The five labeled geodesics are cut into five pairs.
		The black hole microscopic degrees of freedom 
		are whether each pair has aligned arrows (pair 3 here)
		or anti-aligned arrows (pair 1, 2, 4, 5 here).
	} 
	\label{Fig_BH}
\end{figure}

Another concept made clear in the dual picture is 
the black hole microstates,
which turn out to be non-locally encoded
on the horizon and also on the boundary.

In Ref.~\cite{Yan2018arXiv},
we used the increase of ground state entropy
in the bulk
to compute the black hole entropy.
An equivalent definition of black hole entropy
is the entropy from the 
microstates of the black hole \cite{Hawking}.
In the spin picture from the hyperbolic fracton model,
how to identify them is a bit obscure:
the microstate dofs  are not the spins next to the horizon,
since  they are collectively constrained by the non-local symmetry structure, and not independent from each other.

In the dual vertex model,
the microstates of the black hole become clear.
Let us take the black hole in Fig.~\ref{Fig_BH} as an example.
There are five  geodesics cut open by the black hole.
So attached to the horizon are ten  threads,
extending to the boundary.

Let us first consider the original ground states
without the black hole.
From the boundary point of view,
they are those that each pair of 
threads 
aligned in the same direction, 
so that each geodesic has zero subsystem charge.
We define the normalized subsystem charge
\begin{equation}
c_i = \frac{C_i}{2}\mod 2,
\end{equation}
where the $C_i$ denotes the subsystem charge
from the $i$-th pair of bit-threads observed from the boundary.
The ground states then can be expressed collectively as states satisfying
\begin{equation}
(c_1,c_2,c_3,c_4,c_5) = (0, 0, 0, 0, 0).
\end{equation}

After introducing the black hole,
the two bit-threads in each pair become independent.
For the boundary, that means the normalized subsystem charges
for these pairs can be 
\begin{equation}
c_i = 1, \ \text{or } 0.
\end{equation}

The
different black hole microstates
correspond to
different arrays $(c_1,c_2,c_3,c_4,c_5)$.
That is, the dofs living on the horizon are
whether each pair of threads is aligned or not.
Or in more mathematical terms,
the microstates are all the ground states quotient 
the subsystem symmetries from the no-black-hole bulk.
Here we emphasis that
the single bit-threads should  not be viewed as the dofs individually.
This is a critical to identify the correct
black hole microstates:
different states connected by subsystem symmetries
should not be counted,
since they are already included in the 
entropy contribution of ground states without black holes.
This is also reason we use the normalized
subsystem charge $c_i$
instead of the original $C_i$: 
to guarantee that the microstate is
invariant under subsystem symmetries.

A sanity check is to consider the 
``entanglement entropy'' as  half the classical mutual information
between the black hole and the AdS boundary.
As mentioned, the mutual information
is counted by the number of threads 
ending on the horizon on one side and 
the AdS boundary on the other side.
So the entanglement entropy is counted by this number divided by two,
i.e., each pair of thread counts as one dof.
This is consistent with the microstate dof counting.

One interesting implication of the result is that
the black hole dofs are encoded  \textit{non-locally}.
A single thread of a pair
only gives some information of $C_i$
but no information of $c_i$ at all.
Only when both bit-threads
are known can we recover the value of $c_i$.
Thus the black hole microstate information
is non-locally encoded on its horizon
and also the AdS boundary.

\begin{figure}[t]
\centering
\includegraphics[width=0.45\textwidth]{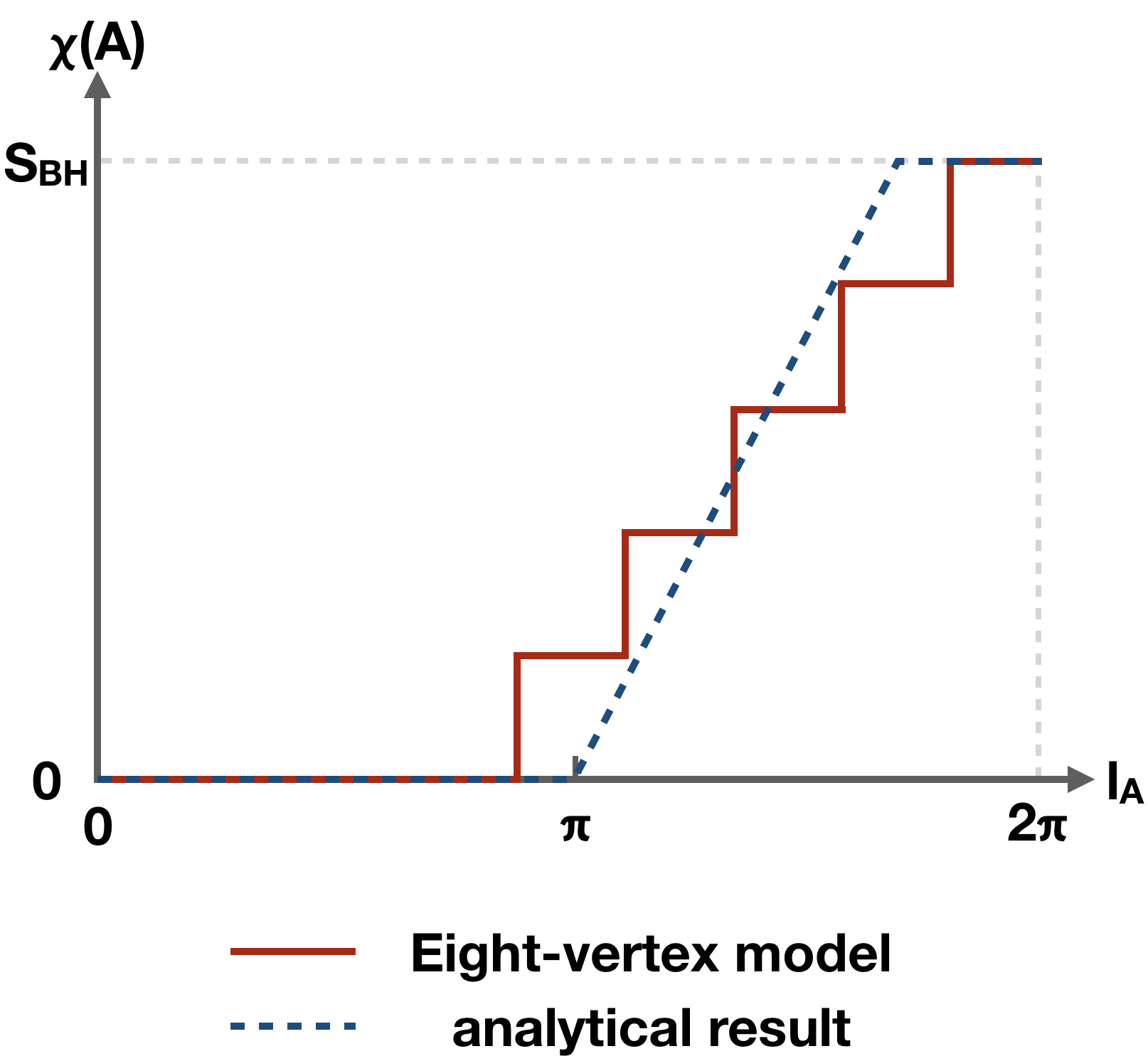}
\caption{
Black hole microstate information 
for observer covering a subregion of the boundary.
Red line: in the eight-vertex model, 
the observer starts to have black hole microstate information
when covering a pair of geodesics cut open by the black hole (Fig.~\ref{Fig_BH}).
Such information is zero until the observer reaches about half the boundary size,
and gradually grows till the observer almost covers the entire boundary.
Blue line:
analytical calculation of the Holevo information 
measuring the microstate distinguishability 
as a function of the boundary subregion area measurable to the observer 
in Ref.~\cite{Bao2017PRD}.
Even though the black hole in the eight-vertex model is very naively defined,
the black hole information recovery behavior looks similar to the analytical results.
} 
	\label{Fig_BH_info}
\end{figure}

Such conclusion agrees with the analysis in Ref.~\cite{Bao2017PRD}, where the authors discussed how much of the AdS boundary subregion needs to be measured 
to distinguish black hole microstates.

In our bit-thread model, as the observer starts to expand the 
observed subregion on the boundary, 
he/she will know the arrow directions of more threads.
But any pair of thread heads from the black hole
is separated by a macroscopic distance,
so starting from zero up to a 
finite subregion,
the observer cannot infer
any information about the black hole microstate.
As the first pair of cut-open threads is included in the observed subregion,
the observer begins to have some information of the black
hole microstates, 
and the amount of  information grows approximately linearly as the subregion
expands. 
Finally when almost covering the full boundary,
the observer can obtain all the information of the black hole microstate.

In Fig.~\ref{Fig_BH_info}, we plot the 
black hole microstate information as a function
of the observed subregion from the eight-vertex model,
as well as the analytical result obtained in Ref.~\cite{Bao2017PRD}.
The behaviors of the two curves qualitatively agree, 
in terms of the zero information segment in the beginning,  the linear growth in the middle
and the final saturation.

\section{Outlook}
\label{Sec_8_outlook}

In this work we discussed in detail the implications
of the dual eight-vertex model equivalent to the original
hyperbolic fracton model.
Despite the equivalence, 
it advances our understanding 
by providing a much clearer picture
of a few aspects of its physics.

The hyperbolic eight-vertex model
becomes a discrete bit-thread model
at zero temperature.
This explains why the fracton model
has the holographic properties demonstrated before.
It is also significant that
we have another  concrete, sophisticated
holographic model -- the bit-thread model --
as a reference frame to
evaluate
the similarity between fracton models 
and the informational-aspects of holography.
It is a very useful guideline
to construct improved holographic fracton models.
For example, fracton bit-threads being discrete
is a major obstacle for holography
at higher order (for disconnected boundary components),
or below the AdS scale (i.e., for regions smaller than the pentagon).
So an improved version should tackle such problems.

The connection between the fracton model
and bit-threads also implies that it might be possible
to establish a concrete duality between linearized gravity
(or theories with linearized diffeomorphism-like gauge symmetry)
and the full-fledged bit-thread model.
It has been pointed out that
rank-2 U(1) gauge theory,
the underlying effective theory of the hyperbolic fracton model (with Higgs mechanism),
is actually the linearized limit of 
certain gravitational theory.
This also gives us some confidence
in constructing more sophisticated
holographic fracton models
to mimic gravity better.

At finite temperature,
utilizing the bit-thread picture
and subsystem charges,
one can establish isometry for a subset of 
low energy states,
and identify the non-locally encoded
black hole microscopic dofs.
It is intriguing 
to ask what will these subsystem charges
become when we work on the 
continuous field theory,
or what is their analogy in gravity.

To explore the relationship between
gravity and fracton states
can be a meaningful program
for condensed matter physics.
A lot is known on how topological 
orders are described 
by gauge theories,
but not much on 
what kind of (beyond) topological order
can arise from gravitational-like theories.
Certain fracton states seem to be such examples
\cite{Pretko2017,gromov19PRL,Slagle2019SciPost},
but the whole picture is vastly unexplored.

If we could discover more gravity-like
many-body systems,
they may also help us 
establish links between gravity
and 
various other toy models of holography,
including the holographic tensor networks
and the bit-thread model.
This work already serves as a primitive 
example of the latter case.
It is also attractive 
to mimic gravity in a laboratory 
using fracton states,
after we understand their relations better.

\section*{Acknowledgment}
We thank  Nic Shannon, Ludovic D. C. Jaubert, Owen Benton, Geet Rakala,
Xiao-Liang Qi, Tadashi Takayanagi and Sugawara Hirotaka
for helpful discussions.
In particular we thank Nic Shannon and Owen Benton for
a careful reading of the manuscript.
HY is supported by 
the Theory of Quantum Matter Unit at Okinawa Institute of Science and Technology,
and
the Japan Society for the Promotion of Science (JSPS)
Research Fellowships for Young Scientists.

\bibliography{reference}

\end{document}